\documentclass[superscriptaddress,letterpaper,aps,twocolumn]{revtex4-1}

\usepackage{graphicx}
\usepackage{amssymb,amsfonts,amsmath,xcolor}
\usepackage{setspace}

\newcommand{\mb}[1]{\mathbf{#1}}

\newcommand{\hmb}[1]{\hat{\mathbf{#1}}}
\newcommand{\nocontentsline}[3]{}
\newcommand{\tocless}[2]{\bgroup\let\addcontentsline=\nocontentsline#1{#2}\egroup}

\DeclareRobustCommand{\gobblefour}[4]{}
\newcommand*{\SkipTocEntry}{\addtocontents{toc}{\gobblefour}}

\DeclareRobustCommand{\gobbleone}[4]{}
\newcommand*{\SkipTocEntryB}{\addtocontents{toc}{\gobbleone}}

\begin{document}

\title{From mechanical folding trajectories to intrinsic energy
  landscapes of biopolymers}

\author{Michael Hinczewski} 

\affiliation{Biophysics Program, Institute for Physical Science and
  Technology, University of Maryland, College Park, MD 20742}

\email{mhincz@umd.edu or thirum@umd.edu} 

\author{J. Christof M. Gebhardt}

\affiliation{Department of Physics, Technical University of
Munich, 85748 Garching, Germany}
\affiliation{Department of Chemistry and Chemical Biology, Harvard University, Cambridge, MA 02138, U.S.A.}

\author{Matthias Rief}
\affiliation{Department of Physics, Technical University of
Munich, 85748 Garching, Germany}

\author{D. Thirumalai}
\affiliation{Biophysics Program, Institute for Physical Science and
  Technology, University of Maryland, College Park, MD 20742}

\begin{abstract}
  In single molecule laser optical tweezer (LOT) pulling experiments a
  protein or RNA is juxtaposed between DNA handles that are attached
  to beads in optical traps.  The LOT generates folding trajectories
  under force in terms of time-dependent changes in the distance
  between the beads. How to construct the full intrinsic folding
  landscape (without the handles and the beads) from the measured time
  series is a major unsolved problem. By using rigorous theoretical
  methods---which account for fluctuations of the DNA handles,
  rotation of the optical beads, variations in applied tension due to
  finite trap stiffness, as well as environmental noise and the
    limited bandwidth of the apparatus---we provide a tractable
  method to derive intrinsic free energy profiles. We validate the
  method by showing that the exactly calculable intrinsic free energy
  profile for a Generalized Rouse Model, which mimics the two-state
  behavior in nucleic acid hairpins, can be accurately extracted from
  simulated time series in a LOT setup regardless of the stiffness of
  the handles.  We next apply the approach to trajectories from coarse
  grained LOT molecular simulations of a coiled-coil protein based on
  the GCN4 leucine zipper, and obtain a free energy landscape that is
  in quantitative agreement with simulations performed without the
  beads and handles. Finally, we extract the intrinsic free energy
  landscape from experimental LOT measurements for the leucine zipper,
  which is independent of the trap parameters.
\end{abstract}

\maketitle

The energy landscape perspective has provided a conceptual framework
to describe how RNA~\cite{Thirum05Biochem} and
proteins~\cite{Onuchic97ARPC,Dill08ARB,Thirum10ARB} fold.  Some of the key theoretical
predictions, such as folding of proteins and
RNA by the kinetic partitioning mechanism~\cite{Guo95Biopolym} and the
diversity of folding routes~\cite{Klimov05JMB}, have been confirmed by
a number of experiments~\cite{Stigler11}. More refined comparisons require 
mapping the full folding landscape of biomolecules, which has been difficult to achieve.  The situation has dramatically changed with advances in
laser optical tweezer (LOT) experiments, which have been used to
obtain free energy profiles as a function of the extension of
biomolecules under
tension~\cite{Woodside06,Woodside06b,Woodside11NAR,Gebhardt10,Stigler11,Elms12}.

\begin{figure}[t]
\centerline{\includegraphics[width=\columnwidth]{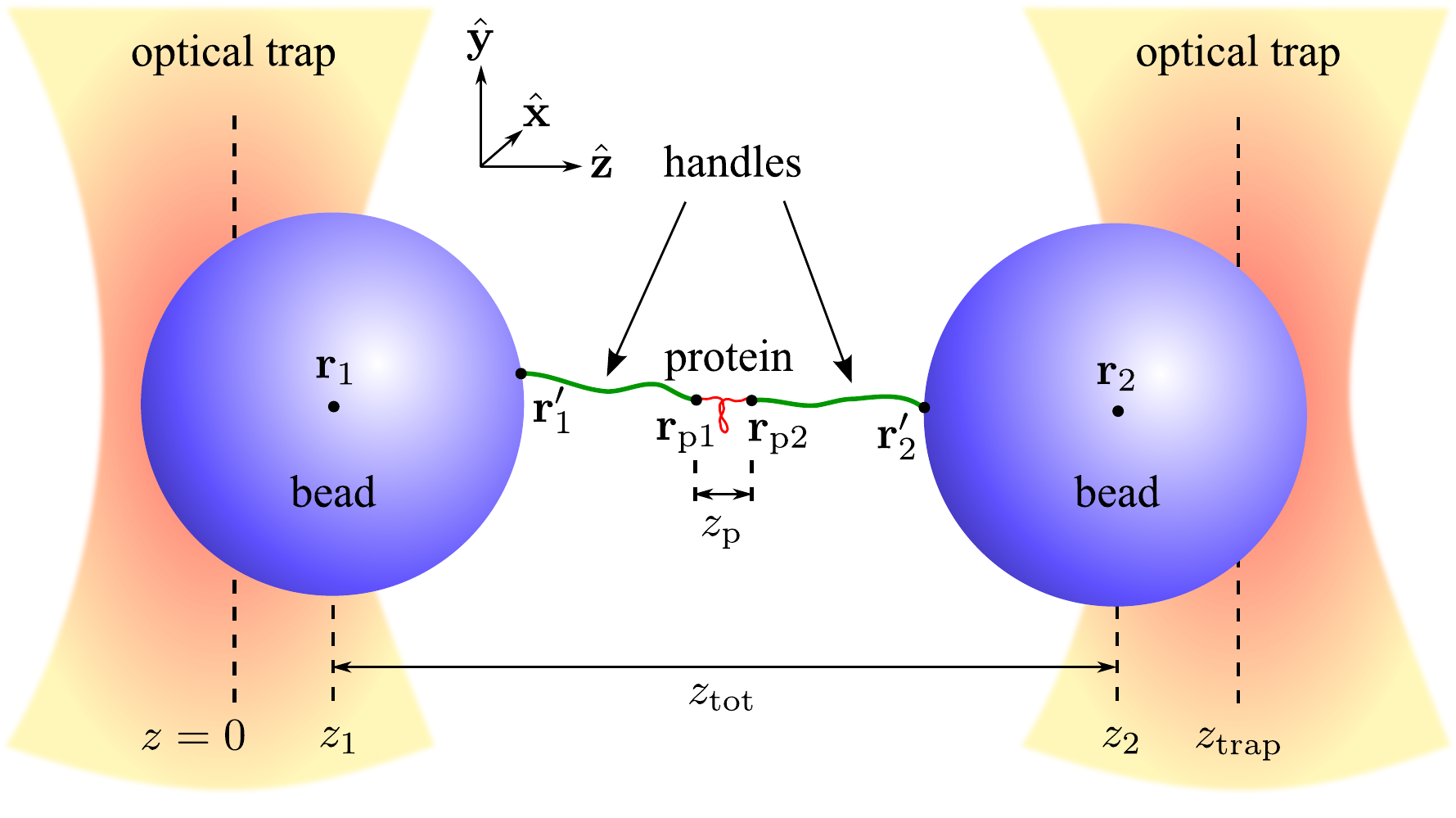}}
\caption{Dual beam optical tweezer  setup for studying the equilibrium
  folding landscape of a single protein molecule under force.}\label{sys}
\end{figure}

The usefulness of the LOT technique, however, hinges on the crucial
assumption that information about the fluctuating biomolecule can be
accurately recovered from the raw experimental data, namely the
time-dependent changes in the positions of the beads in the optical
traps, attached to the biomolecule by double-stranded DNA handles
[Fig.~\ref{sys}].  Thus, we only have access to the intrinsic folding
landscape of the biomolecule (in the absence of handles and beads)
indirectly through the bead-bead separation along the force direction.
Many extraneous factors, such as fluctuations of the
handles~\cite{Hyeon06BJ,Manosas07BJ}, rotation of the beads, and the
varying applied tension due to finite trap stiffness, can severely
distort the intrinsic folding landscape. Moreover, the detectors
  and electronic systems used in the data collection have finite
  response times, leading to filtering of high frequency components in
  the signal~\cite{vonHansen12}.  Ad hoc attempts have been made to
account for handle effects based on experimental estimates of
stretched DNA properties, employing techniques similar to image
  deconvolution~\cite{Woodside06,Gebhardt10,Yu12}.  Theory has been
  used to extract free energy information from nonequilibrium pulling
  experiments~\cite{Hummer10}, and to determine the intrinsic power
  spectrum of protein fluctuations~\cite{Hinczewski10} from LOT data.
However, to date there has been no comprehensive theory to model and
correct for all the systematic instrumental distortions of the
underlying folding landscapes of proteins and RNA.

  A crucial unsolved problem is how can one construct the intrinsic
  free energy profile of a biomolecule using the measured folding
  trajectories in the presence of beads and handles (the total
  separation $z_\text{tot}(t)$ in Fig.~\ref{sys} as a function of time
  $t$).  Here, we solve this problem using a rigorous theoretical
  procedure.  Besides $z_\text{tot}(t)$, the only input needed in our
  theory are  the
  bead radii, the trap strengths and positions, and handle
  characteristics such as the contour length, the persistence length,
  and the elastic stretch modulus.  The output is the intrinsic free
  energy as a function of the biomolecular extension ($z_\text{p}$ in
  Fig.~\ref{sys}) in the constant force ensemble.

  We validate our approach using two  systems: (i) a generalized
  Rouse model (GRM) hairpin~\cite{Hyeon08}, which has an analytically
  solvable double-well  energy landscape under force; this allows
  a direct test of the method; (ii) a double-stranded coiled-coil protein based on
  the yeast transcriptional factor GCN4 leucine zipper domain, whose
  folding landscape was studied using a LOT
  experiment~\cite{Gebhardt10}.  We first use coarse-grained molecular
  simulations to obtain the intrinsic free energy landscape of the
  isolated protein at a constant force.  We then simulate mechanical
  folding trajectories using the full LOT setup, from which we
  quantitatively recover the intrinsic free energy landscape of GCN4,
  thus further establishing the efficacy of our theory.
  Finally, we apply our theory to experimentally generated data, and
  show that we can get reliable estimates for the protein energy
  profile independent of the optical trap parameters.  

\tocless\section{Results}

\tocless\subsection{Theory for constructing the intrinsic protein folding
  landscape from measurements} In a dual beam optical tweezer setup
(Fig.~\ref{sys}) the protein is covalently connected to
double-stranded DNA handles that are attached to glass or polystyrene
beads in two optical traps.  For small displacements of the beads from
the trap centers~\cite{Greenleaf05}, the trap potentials are harmonic,
with strengths $k_x = k_z \equiv k_\text{trap}$ along the lateral
plane, and a weaker axial strength $k_y = \alpha k_\text{trap}$, where
$\alpha < 1$~\cite{Neuman04}.  For simplicity, we take both traps to
have equal strengths, though our method can be generalized to an
asymmetric setup.  The trap centers are separated from each other
along the $\hat{\mb{z}}$ axis, with trap 1 at $z=0$ and trap 2 at
$z=z_\text{trap}$.  As the bead-handle-protein (bhp) system fluctuates
in equilibrium, the positions of the bead centers, $\mb{r}_1(t)$ and
$\mb{r}_2(t)$, vary in time.  We assume that the experimentalist can
collect a time series of the $z$ components of the bead positions,
$z_1(t)$ and $z_2(t)$.  Denote the mean of each time series as $
\bar{z}_{1}$ and $\bar{z}_2$.  We assume that the trap centers are
sufficiently far apart that the whole system is under tension, which
implies that the mean bead displacements are non-zero, $\bar{z}_1 =
z_\text{trap}- \bar{z}_2 = \bar{F}/k_\text{trap} > 0$, where $\bar{F}$
is the mean tension along $\hat{\mb{z}}$.  We focus on the case where
there is no feedback mechanism to maintain a constant force, so the
instantaneous tension in the system changes as the total end-to-end
extension component $z_\text{tot}(t) \equiv z_2(t)-z_1(t)$
[Fig.~\ref{sys}] varies.  Though we choose one particular passive
setup, the theory can be adapted to other types of passive optical
tweezer systems~\cite{Greenleaf05,Woodside06}, where the force is
approximately constant (in which case we could skip the transformation
into the constant-force ensemble described below).  The mean tension
$\bar{F}$, a measure of the overall force scale, can be tuned at the
start of the experiment by making the trap separation $z_\text{trap}$
larger (leading to higher $\bar{F}$) or smaller (leading to lower
$\bar{F}$).  Because $\bar{F} = k_\text{trap} (z_\text{trap} -
\bar{z}_\text{tot})/2$, the precise relationship between
$z_\text{trap}$ and $\bar{F}$ requires knowing the mean total
extension $\bar{z}_\text{tot}$, which depends among other things on
the details of the energy landscape.  Hence, we cannot in general
calculate beforehand what $\bar{F}$ will be for a given
$z_\text{trap}$. However, one of the advantages of our approach is
that we can combine data from different experimental runs (each having
a different $z_\text{trap}$ and $\bar{F}$) to accurately construct the
protein free energy profile.  This combination is carried out through
the weighted histogram analysis method (WHAM)~\cite{Ferrenberg89} (see
Supplementary Information (SI) for details), in a spirit similar
  to earlier work in the context of optical
  tweezers~\cite{Shirts08,Messieres11}.  We first solve the problem of
  obtaining the protein landscape based on a single observed
  trajectory of bead-to-bead separations specified as $z_\text{tot}$
as a function of $t$.

  The key quantity in the construction procedure is ${\cal
    P}_\text{tot}(z_\text{tot})$, the equilibrium probability
  distribution of $z_\text{tot}$ within the external trap potential,
  which can be directly derived from the experimental time series. 
  The imperfect nature of the measured data, due to noise and low-pass
  filtering effects in the recording apparatus, will
  distort ${\cal P}_\text{tot}(z_\text{tot})$, but we have developed a
  technique to model and approximately correct for these issues (see
  Finite Bandwidth Scaling (FBS) in the {\it Methods}).  Once we have
  an experimental estimate for ${\cal P}_\text{tot}(z_\text{tot})$,
  the objective is to find $\tilde{\cal
      P}_\text{p}(z_\text{p};F_0)$, the intrinsic distribution of the
  protein end-to-end extension component $z_\text{p}$ at some {\it
    constant} force $F_0$, whose value we are free to choose. (We will
  use tilde notation to denote probabilities in the constant-force
  ensemble.)  The intrinsic protein free energy profile is
  $\tilde{\cal F}_\text{p}( z_\text{p};F_0) = -k_B T \ln \tilde{\cal
    P}_\text{p}(z_\text{p};F_0)$.  The procedure, obtained from
  rigorous theoretical underpinnings described in detail in the
  SI, consists of two steps:
\vspace{1em}
\begin{enumerate}
\item {\it Transformation into the constant-force ensemble}.  Given
  ${\cal P}_\text{tot}(z_\text{tot})$, we obtain the total system
  end-to-end distribution at a constant $F_0$ using,
\begin{equation}\label{eq:0}
\begin{split}
&\tilde{\cal
    P}_\text{tot}(z_\text{tot};F_0)\\
 &\:= C^{-1} e^{\beta F_0 z_\text{tot} +
  \frac{1}{4}\beta k_\text{trap} ( z_\text{trap} - z_\text{tot})^2} {\cal
    P}_\text{tot}(z_\text{tot}),
\end{split}
\end{equation}
where $\beta = 1/k_B T$ and $C$ is a normalization constant.  The
equation above applies in the case of a single experimental trajectory
at a particular trap separation $z_\text{trap}$.

\item {\it Extraction of the intrinsic protein distribution}.  In the
  constant-force ensemble, $\tilde{\cal P}_\text{tot}
  = \tilde{\cal P}_\text{b} \ast \tilde{\cal P}_\text{h} \ast
  \tilde{\cal P}_\text{p} \ast \tilde{\cal P}_\text{h} \ast
  \tilde{\cal P}_\text{b}$, relates the total end-to-end fluctuations
  $\tilde{\cal P}_\text{tot}(z_\text{tot};F_0)$ to the end-to-end
  distributions for the individual components $\tilde{\cal
    P}_\alpha(z_\alpha;F_0)$, where $\alpha$ denotes bead (b), handle
  (h), or protein (p), and $\ast$ is a 1D convolution
    operator.  For the beads, ``end-to-end'' refers to the
  extension between the bead center and the handle
  attachment point, projected along $\hmb{z}$.  In Fourier space the
  convolution has the form:
\begin{equation}\label{eq:1}
\begin{split}
\tilde{\cal P}_\text{tot}(k;F_0) &= \tilde{\cal P}_\text{b}^2(k;F_0) \tilde{\cal P}^2_\text{h}(k;F_0) \tilde{\cal P}_\text{p}(k;F_0)\\
& \equiv \tilde{\cal P}_\text{bh}(k;F_0) \tilde{\cal P}_\text{p}(k;F_0),
\end{split}
\end{equation}
where $\tilde{\cal P}_\alpha(k;F_0)$ is the Fourier transform of
$\tilde{\cal P}_\alpha(z_\alpha;F_0)$.  Here $\tilde{\cal P}_\text{bh}$,
which is the result of convolving all the bead and handle
distributions, acts as the main point spread function relating the
intrinsic protein distribution $\tilde{\cal P}_\text{p}$ to
$\tilde{\cal P}_\text{tot}$.  Since $\tilde{\cal P}_\text{bh}$ can be
modeled from a theoretical description of the handles and
beads, we can solve for $\tilde{\cal P}_\text{p}$ using
Eq.~\eqref{eq:1} and hence find $\tilde{\cal F}_\text{p}$, the
intrinsic free energy profile of the protein.

\end{enumerate}
The derivation of the procedure (given in the SI, along with technical
aspects of its numerical implementation) shows the conditions under
which the two step method works.  The mathematical approximation
  underlying step 1 becomes exact if either of the following hold:
(i) $k_x = k_y = 0$; (ii) the full 3D total system end-to-end
probability is separable into a product of distributions for
longitudinal ($\hat{\mb{z}}$) and transverse ($\hat{\mb{x}}$,
$\hat{\mb{y}}$) components.  In general, condition (ii) is not
  physically sensible~\cite{Hyeon08}.  However, if
$\bar\rho_\text{tot}$ is the typical length scale describing
transverse fluctuations, then condition (i) is approximately valid
when $\beta k_\text{trap}\bar{\rho}_\text{tot}^2 \ll 1$.  If this
condition breaks down, accurate construction of the intrinsic energy
landscape cannot be performed without knowledge of the transverse
behavior.  However, in the simulation and experimental results below,
the force scales are such that transverse fluctuations are small,
$\bar{\rho}_\text{tot} \sim {\cal O}(1\;\text{nm})$, so to ensure
  condition (i) is met, we require that $k_\text{trap} \ll
  k_BT/\bar{\rho}_\text{tot}^2 = 4.1$ pN/nm at $T= 298$ K.  We use
the experimental value $k_\text{trap} = 0.25\;\text{pN}/\text{nm}$ in
our test cases~\cite{Gebhardt10}, which is well under the upper limit.
In principle, one can choose any $F_0$, the force value of the
constant force ensemble where we carry out the analysis.  In practice,
$F_0$ should be chosen from among the range of forces that is sampled
in equilibrium during the actual experiment, since this will minimize
statistical errors in the final constructed landscape.  For example,
setting $F_0 = \bar{F}$, the mean tension, is a reasonable choice.

Step 2 depends on knowledge of $\tilde{\cal P}_\text{bh}(k;F_0)$, and
thus the individual constant-force distributions of the beads and the
handles in Fourier space.  The point spread function is characterized
by: the bead radius $R_b$, the handle contour length $L$, the handle
persistence length $l_p$, and the handle elastic stretching modulus
$\gamma$.  In $\tilde{\cal P}_\text{h}$ we also include the covalent
linkers which attach the handles to the beads and protein.  If we
model these linkers as short, stiff harmonic springs, we have two
additional parameters: the linker stiffness $\kappa$ and natural
length $\ell$. Using the extensible semiflexible chain as a model
  for the handles, we exploit an exact mapping between this model and
the propagator for the motion of a quantum particle on the surface of
a unit sphere~\cite{Kierfeld04} to calculate the handle Fourier-space
distribution to arbitrary numerical precision.  Together with
analytical results for the bead and linker distributions, we can thus
directly solve for $\tilde{\cal P}_\text{bh}(k;F_0)$.  To verify
  that the analytical model for the point-spread function can
  accurately describe handle/bead fluctuations over a range of forces,
  we have analyzed data from control experiments on a system involving
  only dsDNA handles attached to beads, where ${\cal P}_\text{tot} =
  {\cal P}_\text{bh}$ (SI).  The theory simultaneously fits results
  for several experimental quantities measured on the same system: the
  distributions $\tilde{\cal P}_\text{bh}$ derived from three
  different trap separations, corresponding to mean forces $F_0 =
  9.4-12.7$ pN, and a force-extension curve.  The accuracy of the
  model $\tilde{\cal P}_\text{bh}$ is $\approx 1-3\%$, within the
  experimental error margins.

\tocless\subsection{Robustness of the theory validated by application to an
  exactly soluble model} 
\begin{figure*}[t]
\centerline{\includegraphics*[width=0.98\textwidth]{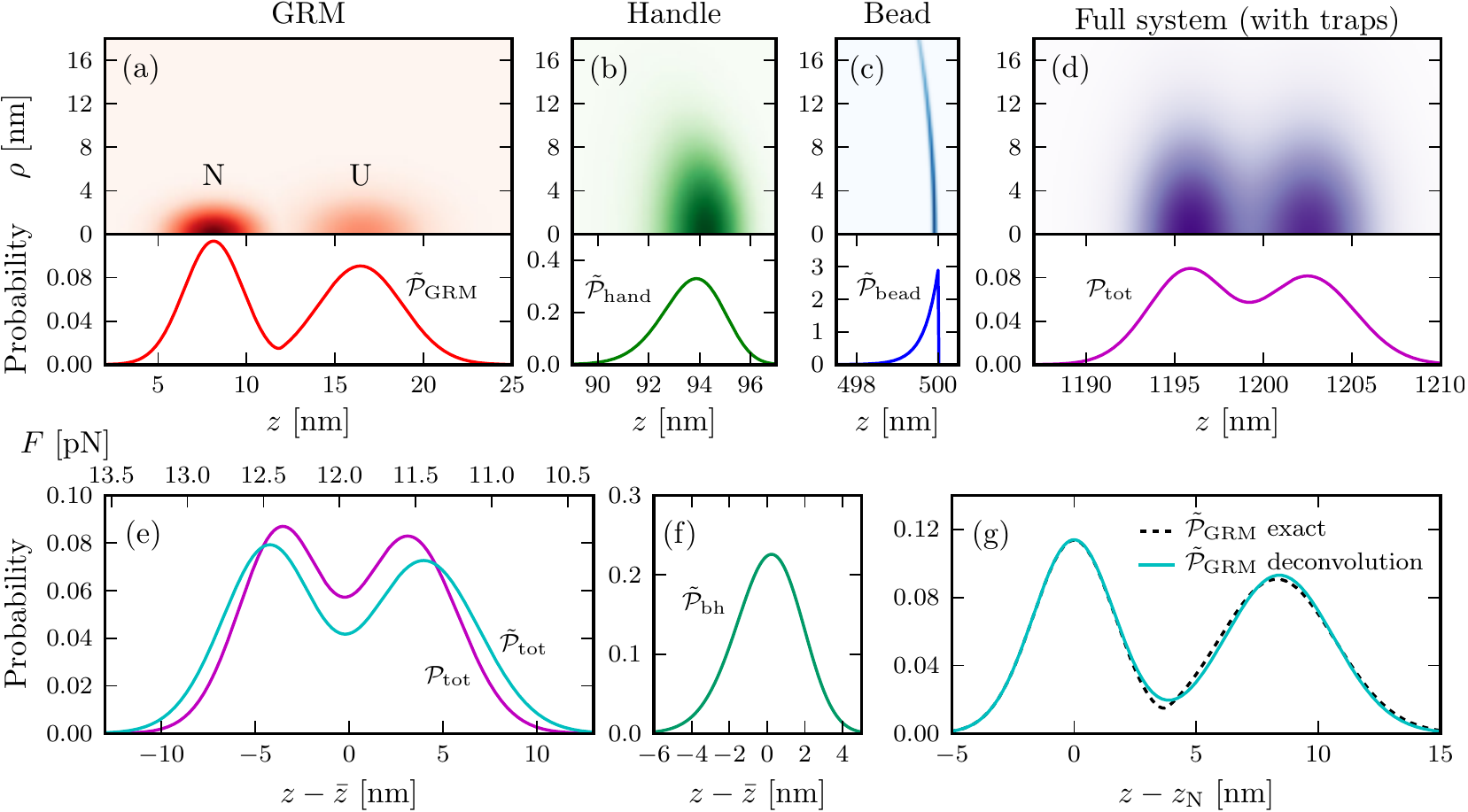}}
\caption{Generalized Rouse model (GRM) hairpin in an optical tweezer
  setup.  The first row shows the exact end-to-end distributions along
  $\hat{\mb{z}}$ for each component type in the system: a) GRM, b)
  dsDNA handle, c) polystyrene bead.  The handle, bead and trap
  parameters are listed in Table~S1 (GRM column).  Upper panels show
  the probabilities projected onto cylindrical coordinates
  $(\rho=\sqrt{x^2+y^2},z)$, while the lower ones show the projection
  onto $z$ alone.  (d) The result for the total system end-to-end
  distribution, ${\cal P}_\text{tot}$, derived by convolving the
  component probabilities and accounting for the optical traps.  (e-g)
  The construction of the original GRM distribution $\tilde{\cal
    P}_\text{GRM}$ starting from ${\cal P}_\text{tot}$.  (e) ${\cal
    P}_\text{tot}$ (purple) and $\tilde{\cal P}_\text{tot}$ (blue) as
  a function of $z$ on the bottom axis, measured relative to
  $\bar{z}$, the average extension for each distribution.  For ${\cal
    P}_\text{tot}$, the upper axis shows the $z$ range translated into
  the corresponding trap forces $F$.  After removing the trap effects,
  $\tilde{\cal P}_\text{tot}$ is the distribution for constant force
  $F_0=11.9$ pN.  (f) $\tilde{\cal P}_\text{bh}$, describing the total
  probability at $F_0$ of fluctuations resulting from both handles and
  the rotation of the beads.  (g) The constructed solution for
  $\tilde{\cal P}_\text{GRM}$ (solid line), obtained by numerically
  inverting the convolution $\tilde{\cal P}_\text{tot} = \tilde{\cal
    P}_\text{bh} \ast \tilde{\cal P}_\text{GRM}$.  The exact
  analytical result for $\tilde{\cal P}_\text{GRM}$ is shown as a
  dashed line. $z_\text{N}$ is the position of the native state (N)
  peak.}\label{grmA}
\end{figure*}

We first apply the theory to a problem for
which the intrinsic free energy profiles at arbitrary force are known
exactly.  The generalized Rouse model (GRM) hairpin (see SI for
details) is a two-state folder whose full 3D equilibrium end-to-end
distributions are analytically solvable.  A representative GRM
distribution $\tilde{\cal P}_\text{GRM}$ at $F_0 = 11.9$ pN is plotted
in Fig.~\ref{grmA}(a).  Since $\tilde{\cal P}_\text{GRM}$ is
cylindrically symmetric, the top panel shows a projection onto the
$(\rho = \sqrt{x^2+y^2},z)$ plane, while the bottom panel shows the
further projection onto the $z$ coordinate.  The two peaks correspond
to the native (N) state at small $z$, and the unfolded (U) state at
large $z$.  In order to model the optical tweezer system, we add
handles and beads to the GRM hairpin, whose probabilities $\tilde{\cal
  P}_\text{h}$ and $\tilde{\cal P}_\text{b}$ (including transverse
fluctuations) are illustrated in Fig.~\ref{grmA}(b) and (c).  The
  full 3D behavior is derived in an analogous manner to the theory
  mentioned above for the 1D Fourier-space distribution $\tilde{\cal
    P}_\text{bh}(k;F_0)$ of the beads/handles; the only difference is
  that the transverse degrees of freedom are not integrated out.  The
3D convolution of the system components, plus the optical trap
contribution, gives the total distribution ${\cal P}_\text{tot}$ in
Fig.~\ref{grmA}(d).  The bead, handle, linker, and trap parameters are
listed in SI Table~S1.  From ${\cal P}_\text{tot}$ one can calculate
the mean total $z$ extension and the mean tension, which in this case
are $\bar{z}_\text{tot} = 1199$ nm, $\bar{F} =
k_\text{trap}(z_\text{trap}-\bar{z}_\text{tot})/2 = 11.9$ pN.

\begin{figure*}[t]
\centerline{\includegraphics*[width=0.98\textwidth]{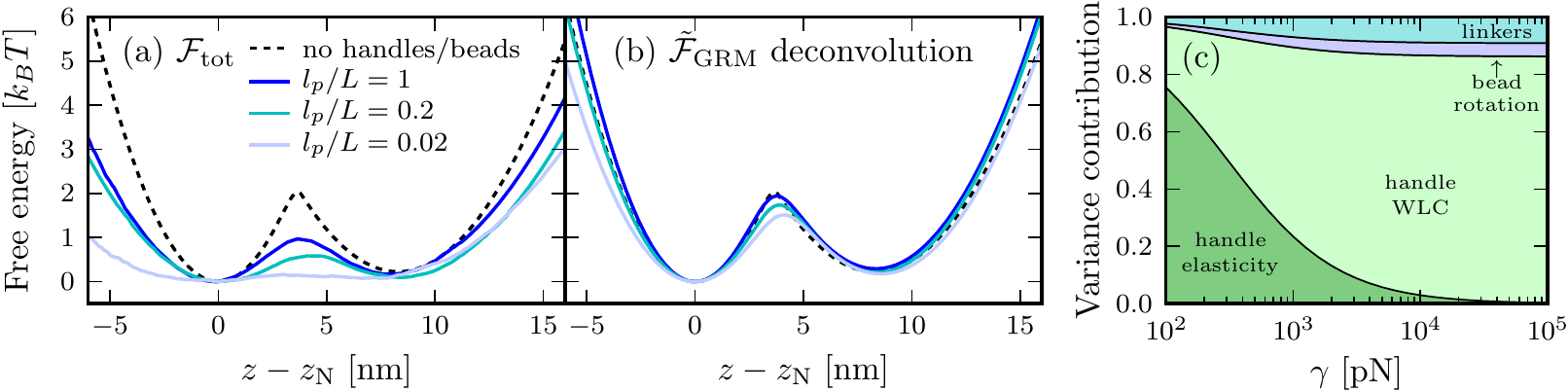}}
\caption{Effects of handle characteristics on the free energy
  profile of the GRM in an LOT setup.
  (a) The total system free energy ${\cal F}_\text{tot} = -k_BT \ln
  {\cal P}_\text{tot}$ for fixed $L=100$ nm, and varying ratios
  $l_p/L$.  All the other parameters are in Table~S1 (GRM
  column).  The exact analytical free energy at $F_0=11.9$ pN  (dashed line) for the
  GRM alone, $\tilde{\cal F}_\text{GRM} = -k_B T \ln \tilde{\cal
    P}_\text{GRM}$, is  shown for comparison.  (b)
  For each ${\cal F}_\text{tot}$ in (a), the construction of
  $\tilde{\cal F}_\text{GRM}$ at $F_0$, together with the exact answer
  (dashed line). (c) For system parameters matching the experiment
  (Table~S1), the variance of the point spread function
  $\tilde{\cal P}_\text{bh}$ broken down into the individual handle,
  bead, and linker contributions.  The fraction for each component is
  shown as a function of varying handle elastic modulus
  $\gamma$.}\label{grmB}
\end{figure*}

The $\mb{\hat{z}}$ probability projection in the bottom panel of (d)
is the information accessible in an experiment, and the computation of
the intrinsic distribution in the bottom panel of (a) is the ultimate
goal of the construction procedure.  Comparing (a) and (d), two
effects of the apparatus are visible: the GRM peaks have been
partially blurred into each other, and the transverse ($\rho$)
fluctuations have been enhanced.  The handles provide the dominant
contribution to both these effects.

Figs.~\ref{grmA}(e) through (g) illustrate the construction procedure
for the GRM optical tweezer system.  Panel (e) corresponds to Step 1,
with a transformation of the distribution ${\cal P}_\text{tot}$ (whose
varying force scale is shown along the top axis) into $\tilde{\cal
  P}_\text{tot}$ at constant force $F_0 = 11.9$ pN.  Step 2 uses the
exact $\tilde{\cal P}_\text{bh}$, shown in real-space in panel (f),
and produces the intrinsic distribution $\tilde{\cal P}_\text{GRM}$,
drawn as a solid line in (g).  The agreement with the exact analytical
result (dashed line) is extremely close, with a median error of $3\%$
over the range shown.  This deviation is due to the approximation
  in Step 1, discussed above, as well as the numerical implementation
  of the deconvolution procedure.

As shown in our previous study~\cite{Hyeon08}, the smaller the ratio
$l_p/L$ for the handles, the more the features of the protein energy
landscape get blurred by the handle fluctuations.  Since the
experimentally measured total distribution always distorts to some
extent the intrinsic protein free energy profile due to the finite
duration and sampling of the system trajectory, more flexible handles
will exacerbate the signal-to-noise problem.  To illustrate this
effect, we performed Brownian dynamics simulations of the GRM in the
optical tweezer setup, with handles modeled as extensible,
semiflexible bead-spring chains (see SI for details).  In
Fig.~\ref{grmB}(a) we compare the free energy ${\cal
  F}_\text{tot} = -k_BT \ln {\cal P}_\text{tot}$ for a fixed $L = 100$
nm and a varying $l_p/L$, derived from the simulation trajectories,
and the exact intrinsic GRM result $\tilde{\cal F}_\text{GRM} = -k_BT
\ln \tilde{\cal P}_\text{GRM}$ at $F_0$.  When the handles are very
flexible, with $l_p/L= 0.02$, the energy barrier between the native
and unfolded states almost entirely disappears in ${\cal
  F}_\text{tot}$, with the noise making the precise barrier shape
difficult to resolve.  Remarkably, even with this extreme level of
distortion, using our theory we still recover a reasonable estimate of
the intrinsic landscape [Fig.~\ref{grmB}(b)].  For each ${\cal
  F}_\text{tot}$ in Fig.~\ref{grmB}(a), panel Fig.~\ref{grmB}(b)
compares the result of the construction procedure and the exact answer
for $\tilde{\cal F}_\text{GRM}$.  Clearly some information is lost as
$l_p/L$ becomes smaller, since the $l_p/L = 0.02$ system does not
yield as accurate a result as the ones with stiffer handles.  However
in all cases the basic features of the exact $\tilde{\cal
  F}_\text{GRM}$ are reproduced.  Thus, the theoretical-based method
works remarkably well over a wide range of handle parameters.  This
conclusion is generally valid even when other parameters are varied
(see Fig.~S3 in the SI for tests at various $F_0$ and
$k_\text{trap}$).  The excellent agreement
between the constructed and intrinsic free energy profiles for the
exactly solvable GRM hairpin over a wide range of handle and trap
experimental variables establishes the robustness of the theory.

\begin{figure*}[t]
\centerline{\includegraphics*[width=0.98\textwidth]{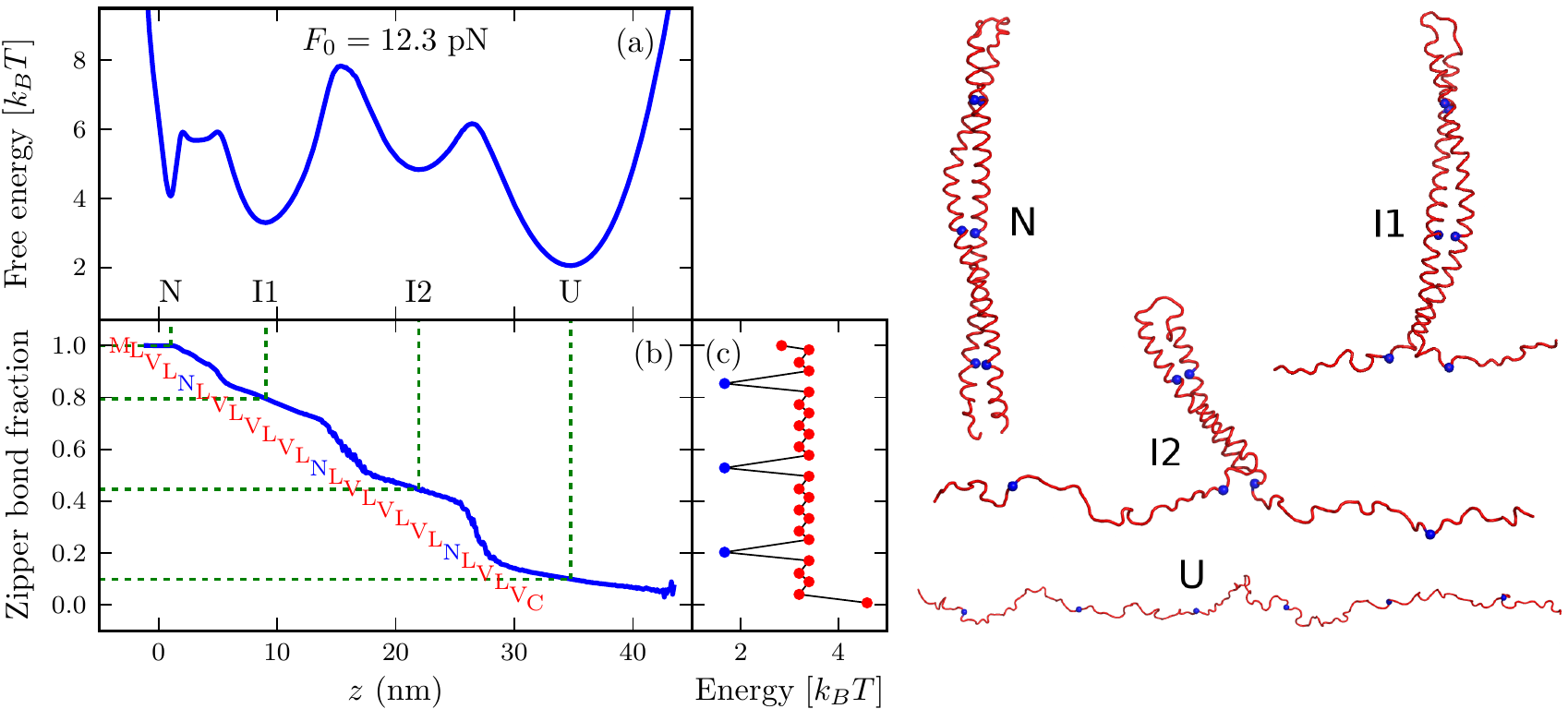}}
  \caption{Intrinsic characteristics of the LZ26 leucine zipper at
    constant $F_0$, derived from SOP simulations in the absence of
    handles/beads.  (a) LZ26 free energy $\tilde{\cal F}_\text{p}$ at
    $F_0 = 12.3$ pN vs. end-to-end extension $z$.  Representative
    protein configurations from the four wells (N, I1, I2, U) are
    shown on the right, with asparagine residues colored blue. (b) The
    average fraction of native contacts between the two alpha-helical
    strands of LZ26 (the ``zipper bonds'') as a function of $z$.
    Listed to the left of the curve are the $a$ and $d$ residues in
    the heptads making up the amino acid sequence for each LZ26
    strand, placed according to their position along the zipper.
    Asparagines (N) are highlighted in blue.  (c) For the residues
    listed in (b), the residue contact energies used in the SOP
    simulation (rescaled BT~\cite{Betancourt99} values).}\label{LZ26}
\end{figure*}

\begin{figure*}
\centerline{\includegraphics*[width=0.98\textwidth]{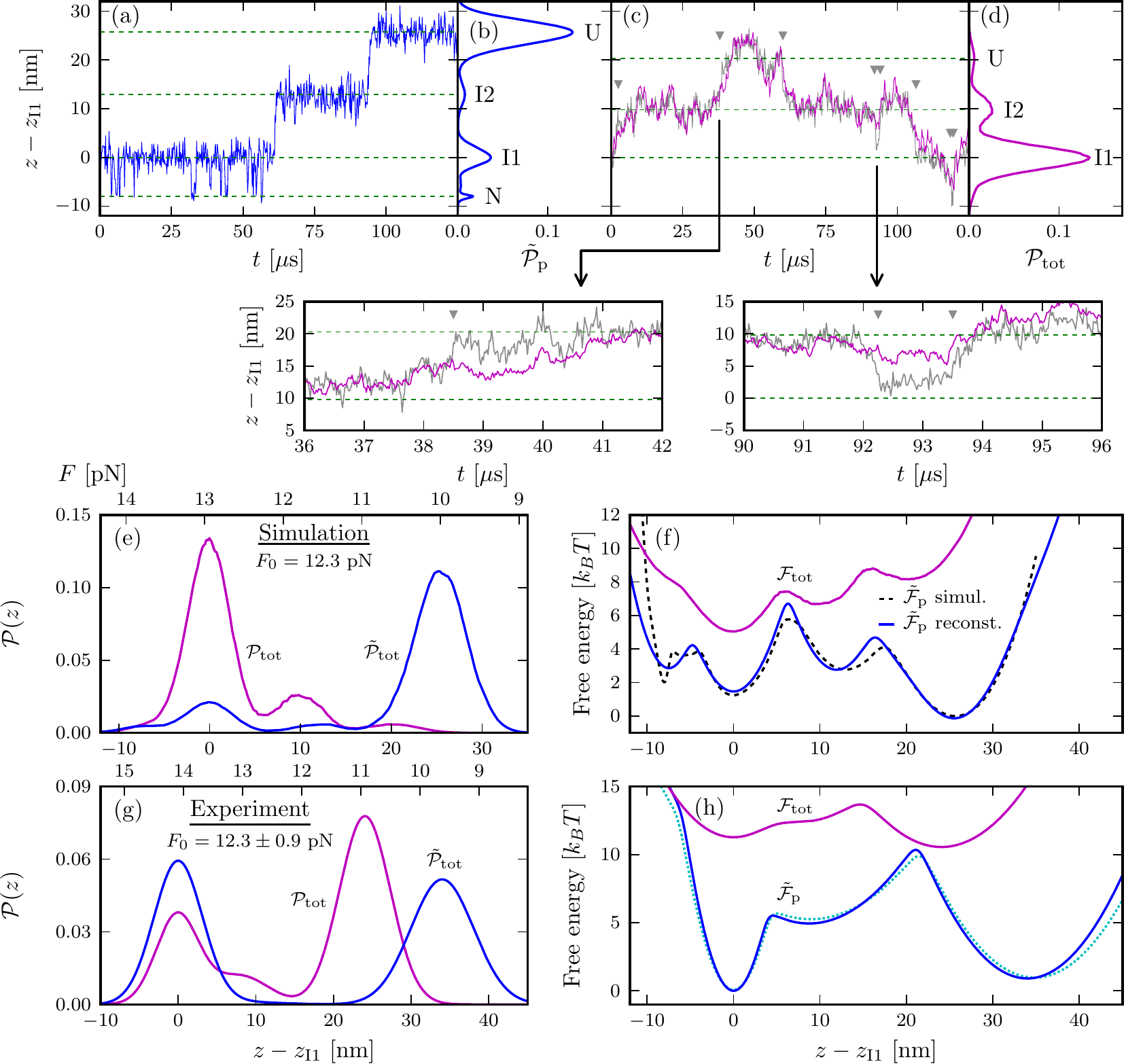}}
  \caption{(a,b) A trajectory fragment and the probability
    distribution $\tilde{\cal P}_\text{p}$ from SOP simulations of the
    LZ26 leucine zipper at constant force $F_0 = 12.3$ pN in the
    absence of handles/beads. (c,d) A trajectory fragment and the
    total system distribution ${\cal P}_\text{tot}$ at $z_\text{trap}
    = 503$ nm.  Panel (c) shows both the total extension
    $z_\text{tot}(t)$ (purple) and the protein extension
    $z_\text{p}(t)$ (gray).  Triangles mark times when the protein
    makes a transition between states, and the arrows point to two
    enlarged portions of the trajectories.  In all cases the $z$-axis
    origin is $z_\text{I1}$, the peak location of the I1 intermediate
    state.  (e-g) Leucine zipper free energy profiles extracted from
    time series (third row = simulation, fourth row = experiment). The
    first column shows the total system end-to-end distribution ${\cal
      P}_\text{tot}$, and the corresponding $\tilde{\cal
      P}_\text{tot}$ at constant force $F_0 = 12.3$ pN.  In the
    experimental case $F_0 = 12.3 \pm 0.9$ pN is the mid-point force
    at which the I1 and U states are equally likely.  For ${\cal
      P}_\text{tot}$, $z_\text{trap} = 503$ nm (simulation), $1553\pm
    1$ nm (experiment).  Force scales at the top are the range of trap
    forces for ${\cal P}_\text{tot}$.  The second column shows the
    computed intrinsic protein free energy profiles $\tilde{\cal
      F}_\text{p}$, compared to the total system profile, ${\cal
      F}_\text{tot}$ (shifted upwards for clarity).  (f) SOP
    simulations for the protein alone at constant $F_0$ provide a
    reference landscape, drawn as a dashed line.  (h) The dashed curve
    is the reconstructed $\tilde{\cal F}_\text{p}$ at the mid-point
    force $F_0 = 12.1 \pm 0.9$ pN, from a second, independent
    experimental trajectory, with $z_\text{trap} = 1547 \pm 1$ nm.
    The $\tilde{\cal F}_\text{p}$ curves have a median uncertainty of
    $0.4$ $k_BT$ over the plotted range (see SI for error analysis).
  }\label{deconv}
\end{figure*}

\vspace{1em}

\tocless\subsection{Intrinsic folding landscape of a simulated leucine
  zipper} To demonstrate that the theory can be used to produce
equilibrium intrinsic free energy profiles with multiple states from
mechanical folding trajectories, we performed simulations of a protein
in an optical tweezer setup.  The simulations were designed to mirror
the single-molecule experiment reported in Ref.~\cite{Gebhardt10}, and
to this end we studied a coiled-coil, LZ26~\cite{Bornschlogl06}, based
on three repeats of the leucine zipper domain from the yeast
transcriptional factor GCN4~\cite{OShea91} (see {\it Methods}). The
simple linear unzipping of the two strands of LZ26 allows us to map
the end-to-end extension to the protein configuration.  Furthermore,
the energy heterogeneity of the native bonds that form the ``teeth''
of the zipper leads to a non-trivial folding landscape with at least
two intermediate states~\cite{Bornschlogl06,Bornschlogl08,Gebhardt10}.
The more complex landscape of LZ26 thus provides an additional
stringent test of the proposed theory.

The native (N) structure of LZ26 is illustrated on the right in
Fig.~\ref{LZ26} (from a simulation snapshot), with the two
alpha-helical strands running from N-terminus at the bottom to
C-terminus at the top.  In the experiment a handle is attached to the
N-terminus of each strand, and this is where the strands begin to
unzip under applied force.  To prevent complete strand separation, the
C-termini are cross-linked through a disulfide bridge between two
cysteine residues.  Each alpha-helix coil consists of a series of
seven-residue heptad repeats, with positions labeled a through g.  For
the leucine zipper the a and d positions are the ``teeth'', consisting
of mostly hydrophobic residues (valine and leucine) which have strong
non-covalent interactions with their counterparts on the other strand.
The exceptions to the hydrophobic pattern are the three hydrophilic
asparagine residues in $a$ positions on each strand (marked in blue in
the structure snapshots on the right of Fig.~\ref{LZ26}).  As has been
seen experimentally~\cite{Bornschlogl06,Gebhardt10} (and shown below
through simulations), the weaker interaction of these asparagine pairs
is crucial in determining the properties of the intermediate folding
states, a point we will return to in more detail in the Discussion.

In analyzing the LZ26 leucine zipper system, we performed
coarse-grained simulations using the Self-Organized Polymer (SOP)
model~\cite{Hyeon06} (full details in the SI, with selected parameters
summarized in Table~S1).  The intrinsic free energy profile
$\tilde{\cal F}_\text{p} = -k_B T \ln \tilde{\cal P}_\text{p}$ at $F_0
= 12.3$ pN is shown in Fig.~\ref{LZ26}(a).  The four prominent wells
in $\tilde{\cal F}_\text{p}$ as a function of $z_\text{p}$ correspond
to four stages in the progressive unzipping of LZ26.  At $F_0=12.3$ pN
all the states are populated, and the system fluctuates in equilibrium
between the wells.  The transition barrier between N and I1 exhibits a
shallow dip that may correspond to an additional, very transiently
populated intermediate.  Since this dip is much smaller than $k_B T$,
we do not count it as a distinct state.

Like in the GRM example, adding the optical tweezer apparatus to the
SOP simulation significantly distorts the measured probability
distributions.  In the first row of Fig.~\ref{deconv} sample
simulation trajectory fragments are shown both for the protein-only
case [Fig.~\ref{deconv}(a)] at constant force $F_0 = 12.3$ pN, and
within the full optical tweezer system [Fig.~\ref{deconv}(c)] with
$z_\text{trap} = 503$ nm.  For the
  latter case we plot both $z_\text{tot}(t)$ (purple) and
  $z_\text{p}(t)$ (gray), allowing us to see how the bead separation
  tracks changes in the protein extension.  The probability
distributions $\tilde{\cal P}_\text{p}$ and ${\cal P}_\text{tot}$ are
plotted in Fig.~\ref{deconv}(b) and (d) respectively.  In
Fig.~\ref{deconv}(e), the distribution ${\cal P}_\text{tot}$ within
the optical tweezer system is plotted for $z_\text{trap} = 503$ nm.
Though we only illustrate this particular $z_\text{trap}$ value,
$\approx 260$ trajectories are generated at different $z_\text{trap}$
and combined together using WHAM~\cite{Ferrenberg89} (see SI) to
produce a single $\tilde{\cal P}_\text{tot}$ at a constant force $F_0
= 12.3$ pN [Fig.~\ref{deconv}(e)].  We can then use our theoretical
method to recover the protein free energy $\tilde{\cal F}_\text{p}$
[Fig.~\ref{deconv}(f)].  Despite numerical errors due to limited
statistical sampling (both in the protein-only and total system runs),
there is remarkable agreement between the constructed result and
$\tilde{\cal F}_\text{p}$ derived from protein-only simulations.  This
is particularly striking given that the total system free energy
${\cal F}_\text{tot}(z_\text{tot})=-k_B T \ln {\cal
  P}_\text{tot}(z_\text{tot})$, plotted for comparison in panels (f),
shows how severely the handles/beads blur the energy landscape,
reducing the energy barriers to a degree that the N state is difficult
to resolve.  The signature of N in ${\cal F}_\text{tot}(z_\text{tot})$
is a slight change in the curvature at higher energies on the left of
the I1 well.  However despite this, we still recover a basin of
attraction representing the N state in the constructed $\tilde{\cal
  F}_\text{p}$.  Overall, the results in (f) show that our theory can
accurately produce the intrinsic free energy profiles using only the
simulated folding trajectories as input, thus proving a
self-consistency check of the method for a system with multiple
intermediates.

\tocless\subsection{Folding landscape of the leucine zipper from experimental
  trajectories} As a final test of the
efficacy of the theory we used the experimental time series
data~\cite{Gebhardt10} to obtain $\tilde{\cal F}_\text{p}$.  The data
consists of two independent runs with the LZ26 leucine zipper, using
the same handle/bead parameters for each run (see Table~S1), but at
different trap separations $z_\text{trap}$.  We project the
deconvolved landscape from each trajectory onto the mid-point force
$F_0$ where the two most populated states (I1 and U) have equal
probabilities in $\tilde{\cal P}_\text{p}$.  The values of $F_0$
derived from the two runs are the same within error bounds: $12.3 \pm
0.9$ and $12.1 \pm 0.9$ pN.  The detailed deconvolution steps are
shown for one run in the last row of Fig.~\ref{deconv}, and the final
result, the intrinsic free energy profile $\tilde{\cal F}_\text{p}$,
is shown for both runs in Fig.~\ref{deconv}(h) (solid and dotted blue
curves respectively).  Accounting for error due to finite trajectory
length and uncertainties in the apparatus parameters, the median total
uncertainty in each of the reconstructed landscapes is about $0.4$
$k_BT$ in the $z$ range shown (see SI for full error analysis).  The
landscapes from the two independent runs have a median difference of
$0.3$ $k_B T$, and hence the method gives consistent results between
runs, up to a small experimental uncertainty, an important test of its
practical utility.  The reproducibility of $\tilde{\cal F}_\text{p}$
is a testament to the stability of the dual optical tweezer setup,
allowing us to sample extensively from the energy landscape: each
trajectory lasted for more than 100 s, and thus collected $\sim {\cal
  O}(10^2-10^5)$ of the various types of transitions between protein
states (the slowest transition, $\text{U}\to\text{I2}$, occurred on
time scales of $0.4 - 0.6$ s).

Comparison between the experimental $\tilde{\cal F}_\text{p}$ in panel
(h) and the simulation result in (f) reveals a notable difference: the
landscape constructed using the experimental data does not have four
identifiable basins. The N state may not be discernible in the
experiment because of the limited resolution of the apparatus
(see below).  The spacing between the I1 and I2 wells is similar
  in the simulation and experiment ($\approx 9-13$ nm), but that
  between I2 and U is $\approx 13$ nm in the simulation versus 25 nm
  in the experiment.  This is likely due to a larger helix content in
the unfolded state for the simulation case.

\vspace{2em}

\tocless\section{Discussion}

\tocless\subsection{Origins of the variance in the point spread function} Our
 theory for the point spread function $\tilde
P_\text{bh}$ can be used to understand the interplay of physical
effects that relate the intrinsic protein distribution to the total
system.  To quantify the various contributions to $\tilde
P_\text{bh}$, we calculated its variance.  Since variances of
probability distributions combine additively upon convolution, we 
break down the variance of $\tilde P_\text{bh}$ into the individual
bead, handle, and linker contributions.  Fig.~\ref{grmB}(c) shows the
fraction of the variance associated with each component as a function
of the handle elastic stretching modulus $\gamma$ at $F_0=12.3$ pN,
with $R_b = 500$ nm, $L = 188$ nm, $l_p = 20$ nm (the approximate
experimental parameters from Ref.~\cite{Gebhardt10}).  For any given
value of $\gamma$, the height of each of the four colored slices
represents four fractions.  Though not directly measured in
Ref.~\cite{Gebhardt10}, we have assumed $\kappa = 200$
kcal/mol$\cdot$nm$^2$, $\ell = 1.5$ nm for the linkers.  The handle
contribution is itself broken down into  the ``elastic''
part, defined as the extra variance due to the finite stretching
modulus $\gamma$, compared to an inextensible ($\gamma \to \infty$)
worm-like chain (WLC), and the remainder, which we call the WLC part.
For the case of Ref.~\cite{Gebhardt10}, $\gamma = 400$ pN.  Since the
length extension relative to the WLC result is $\approx F_0/\gamma$,
we expect finite handle extensibility to play a small role.
However, the elastic contribution to the total $\tilde{\cal
  P}_\text{bh}$ variance at this $\gamma$ is 43\%, comparable to the
WLC contribution of 48\%.  Hence, in predicting
$\tilde{\cal P}_\text{bh}$  correctly it is important to account for both the
bending rigidity and elasticity of the handles, which are exactly
modeled in our approach.

\vspace{1em}

\tocless\subsection{Nature and location of the intermediate states in the
  leucine zipper energy landscape} The folding landscape of LZ26 is
apparently closely related to the pattern of residue-residue contact
energies between the two strands of the
zipper~\cite{Bornschlogl06,Bornschlogl08,Gebhardt10}.   SOP
simulations  give us a detailed picture of this relationship.
The average fraction of intact inter-strand (``zipper'') bonds
vs. extension $z$,  in Fig.~\ref{LZ26}(b)  is a
monotonic curve, starting with the fully closed structure on top (N
state, bond fraction near 1) to the fully open structure at the bottom
(U state, bond fraction near 0).  Listed along this curve are the
individual residues at the $a$ and $d$ positions of the heptads in the
sequence.  Several features stand out: the transition barriers between
the states show a steeper rate of zipper bond unraveling compared to
the well regions.  The change of slope from steep to more gradual
descent occurs near the location of the asparagine residues in the
sequence, and the the well minima of I1, I2, and U occur one or two
residues after the asparagines.  The correlation between well minima
locations and asparagines agrees with the experimental
landscape~\cite{Gebhardt10}, underscoring the importance of the weak,
hydrophilic asparagine bonds that interrupt the hydrophobic
valine/leucine pattern at the a/d positions.  The sequence of rescaled
BT~\cite{Betancourt99} energies used for the a/d native contacts is
plotted in Fig.~\ref{LZ26}(c).  The a/d bonds are all $>2.8$ $k_BT$,
except for the asparagines, which are less stable at $1.7$ $k_B T$.

\vspace{1em}

\tocless\subsection{Instrumental noise filtering, and the limits
    of the theoretical approach} The difference in the number of wells
  in the simulation and experimental free energy landscapes of the
  leucine zipper is related to  finite time and spatial
  resolution.  The measured time series is subject to noise
  (environmental vibrations of the optical elements, detector shot
  noise), as well as low-pass filtering due to ``parasitic'' effects
  in the photodiodes and the nature of the electronic amplification
  circuits~\cite{vonHansen12}.  The standard experimental protocol
  often involves additional low-pass filtering as a way of removing
  noise and smoothing trajectories: for the leucine zipper every five
  data points (originally recorded at 10 $\mu$s intervals) are
  averaged together during collection to give a time step of 50
  $\mu$s~\cite{Gebhardt10}; in other cases similar effects are
  achieved using Bessel filters~\cite{Yu12}.  Noise broadens
  the measured distribution of bead separations, while low-pass
  filtering narrows it.  We developed the FBS technique ({\em Methods}
  and SI), based on the details of the specific apparatus used in the
  experiment, to estimate and correct for the distortions.  For our
  system, the FBS theory provides an excellent description, as we have
  verified in tests using both numerical simulations and experimental
  data (with and without the additional filtering).

  However the FBS theory can only apply corrections to peaks
  (i.e. distinct protein states) that we observe in the measured
  probability distributions.  There is the possibility of protein
  states leaving no discernible signature in the recorded
  distribution.  The N state in the leucine
  zipper is only connected to the I1 state in the folding
  pathway.  In the simulations, where the N state is directly
  observed, it has short mean lifetimes ($\lesssim 6$ $\mu$s in the
  studied force range), and the $\text{N}\leftrightarrow\text{I1}$
  change involves the shortest mean extension difference ($\approx 8$
  nm) among all the transitions.  If the N state in the actual protein
  has similar properties, it could be impossible to resolve it in the
  experimental data for two different reasons: (i) Regardless of any
  additional filtering, the intrinsic low-pass characteristics of the
  apparatus filter out states with very short lifetimes.
   For our LOT setup, the
  effective low-pass filter time-scale for the detectors/electronics
  is $\tau_f \approx 7$ $\mu$s (SI), which is at the cutting
  edge of current technology.  Thus, states with
  lifetimes $\lesssim \tau_f$ will not appear as distinct peaks in the
  measured distribution.  (ii) Independent of the filtering issues in
  detection/recording, environmental background noise in the time
  series also poses a problem, particularly since we measure bead
  displacements, and these have signal amplitudes at high frequencies
  that are generally attenuated compared to the intrinsic amplitudes
  of the protein conformational changes.  The reason for this is that
  the beads have much larger hydrodynamic drag than dsDNA handles or
  proteins, and their characteristic relaxation times $\tau_r$ in the
  optical traps may be comparable to or larger than the lifetime of a
  particular protein state.  The bead cannot fully respond to force
  changes on time scales shorter than its relaxation
  time~\cite{Manosas07BJ}.  For example, $\tau_r \approx 20$ $\mu$s in
  the leucine zipper experiment.  If the lifetime of the N state
  at a particular force is much smaller than $\tau_r$, protein
  transitions from I1$\to$N$\to$I1 will generally occur before the
  bead can relax into the N state equilibrium position.  If the bead
  displacements associated with these transitions are smaller than the
  noise amplitude in the time series, the entire excursion to the N
  state will be lost to the noise.

We can illustrate the finite response time of the bead using
simulations where resolution is not limited by
noise or apparatus filtering, allowing us to illustrate the
relationship between $z_\text{tot}(t)$ and $z_\text{p}(t)$, compared
in two different trajectory fragments in Fig.~\ref{deconv}(c).
Triangles in the figure indicate times where the protein makes a
transition between states.  Changes in protein extension during these
transitions are very rapid, and the bead generally mirrors these
changes with a small time lag, as seen in the enlarged
trajectory interval at $t=36-42$ $\mu$s.  When the protein makes
sharp, extremely brief excursions (like a visit to the N state from I1
in the enlarged interval $t=90-96$ $\mu$s), the corresponding changes
in bead separation are smaller and much less well-defined.  In the
presence of noise, such tiny changes would be obscured.

Thus, we surmise that the N state is not observable due to some
combination of apparatus filtering, noise, and finite bead response
time.  Hence, the theory applied to the experimental data produces a
landscape with only I1, I2, and U wells, as opposed to the four wells
produced from the simulation data.  Our labeling of the basins in the
landscape agrees with the earlier state
identification~\cite{Gebhardt10}, and provides an explanation for why
the N state was not resolved.

\vspace{2em}

\tocless\section{Conclusions}

Extraction of the energy landscape of biomolecules using LOT data is
complicated because accurate analysis depends on correcting for
distortions due to system components on the measured result.  We have
solved this problem completely by developing a theoretically-based
construction method that accounts for these factors.  Through an array
of tests involving an analytically solvable hairpin model,
coarse-grained protein simulations, and experimental data, we have
demonstrated the robustness of the technique in a range of realistic
scenarios.  The method works for arbitrarily complicated landscapes,
as demonstrated by the analysis of the leucine zipper experimental
data, producing consistent results when the same protein is studied
under different force scales.

\vspace{2em}

\tocless\section{Materials and Methods}

  \tocless\subsection{Finite Bandwidth Scaling (FBS)} Probability
    distributions derived from experimental time series of bead-bead
    separations are corrupted by 
    noise, low-pass filtering due to the apparatus, and in some cases
    additional filtering due to the data processing protocol.  We 
    developed FBS theory to model and correct for these effects (see
    SI for details), using information encoded in time series
    autocorrelations, together with earlier spectral characterization
    of the dual trap optical tweezer detector and electronic
    systems~\cite{vonHansen12}.  All the experimental distributions
    ${\cal P}_\text{tot}$ in the main text were first processed by
    FBS.

\vspace{1em}

  \tocless\subsection{Leucine zipper} We use a variant of the coarse-grained
  self-organized polymer (SOP) model~\cite{Hyeon06,Mickler07PNAS},
  where each of the 176 residues in LZ26 is represented by a bead
  centered at the $C_\alpha$ position (see SI for details.)  The
  $\alpha$-helical secondary structure is stabilized by interactions
  which mimic $(i,i+4)$ hydrogen bonding~\cite{Denesyuk11}.  We use
  residue-dependent energies for tertiary
  interactions~\cite{Betancourt99}.

\vspace{1em}

\tocless\subsection{Simulations} We simulate (see SI for details)
trajectories for both the protein alone and the full optical tweezer
setup using an overdamped Brownian dynamics (BD)
algorithm~\cite{Ermak78}. The handles used in the LOT setup
[Fig.~\ref{sys}] are modeled as semiflexible chains.  

\vspace{1em}

\SkipTocEntry\begin{acknowledgments}
M.H. was a Ruth L. Kirschstein National Research Service postdoctoral
fellow, supported by a grant from the National Institute of General
Medical Sciences (1 F32 GM 97756-1). D.T. was supported by a grant
from the National Institutes of Health (GM 089685). Part of the work
was done while D.T. was in TUM as a senior Humboldt Fellow.
\end{acknowledgments}

\SkipTocEntry

\newpage

\begin{widetext}

\renewcommand{\theequation}{S\arabic{equation}}
\renewcommand{\thefigure}{S\arabic{figure}}
\renewcommand{\thetable}{S\arabic{table}}
\setcounter{equation}{0}
\setcounter{figure}{0}
\setcounter{section}{0}

\baselineskip=20.74pt
\begin{center}
{\Large\bf Supplementary information for ``From mechanical folding
  trajectories to intrinsic energy landscapes of biopolymers''}\\[1em]
\baselineskip=17.28pt
{\large Michael Hinczewski, J. Christof M. Gebhardt, Matthias Rief, and D. Thirumalai}
\end{center}
\baselineskip=14.4pt

\SkipTocEntryB\listoftables
\tableofcontents

\newpage


\begin{table}[h]
\caption[GRM, SOP, and experimental parameters]{Parameters used in the GRM model and SOP simulation of the LZ26
  leucine zipper, together with the
  corresponding quantities from two experimental
  runs~\cite{Gebhardt10}}\label{tab1}
\vspace{0.5em}
\begin{minipage}{\textwidth}
\centering
\begin{tabular}{|l|c|c|c|c|}
\hline
Parameter & GRM & SOP Simulation & Experiment\\
\hline
Bead: $R_b$ (nm) & 500 & 100  & 500$\pm$25\\
Trap: $z_\text{trap}$ (nm) & 1294 & 483 -- 543 &  1553$\pm$1, 1547$\pm$1\footnote{different separations correspond to two folding trajectories}\\
Trap: $k_\text{trap}$ (pN/nm) & 0.25 & 0.25 & 0.25$\pm$0.03, 0.27$\pm$0.03\footnote{left, right trap strengths}\\
Trap: $\alpha$ & 1/3 & 1/3 & \text{unknown}\\
Handle: $L$ (nm) & 100 & 100 & 188$\pm$2 \\
Handle: $l_p$ (nm) & 20 & 20 & 20$\pm$2\\
Handle: $\gamma$ (pN) & 2780 & 2780 & 400$\pm$40 \\
Linker: $\kappa$ (kcal/mol$\cdot$nm$^2$) & 200 & 200 & 200\footnote{linker characteristics are assumed for the experimental case} \\
Linker: $\ell$ (nm) & 1.5 & 1.5 & 1.5 \\ 
\hline
\end{tabular}

\renewcommand{\footnoterule}{}
\end{minipage}
\end{table}

\section{Theory for free energy construction from mechanical folding time series}

\subsection{Optical trap Hamiltonian} We begin with the Hamiltonian for the
beads in the traps (Fig.~1 in the main text), which allows us to
introduce the relevant variables of the system.  If the displacements
of the beads from the trap centers are small ($< 100$ nm for a laser
of 1064 nm wavelength and bead radii $\sim {\cal
  O}(100\;\text{nm})$~\cite{Greenleaf05}), the trap Hamiltonian can be
approximately written as:
\begin{equation}\label{eq1}
{\cal H}_\text{trap}(\mb{r}_1,\mb{r}_2) = \frac{1}{2}k_x (x_1^2+x_2^2) + \frac{1}{2}k_y (y_1^2+y_2^2) + \frac{1}{2}k_z \left[z_1^2 + (z_\text{trap} - z_2)^2\right],
\end{equation}
where $\mb{r}_i = (x_i,y_i,z_i)$ is the position of the $i$th bead
center, $k_x$, $k_y$, $k_z$ are the trap strengths along each
coordinate direction, and the two traps are positioned at $z=0$ and
$z=z_\text{trap}$ respectively.  Given the cylindrical symmetry of the
optical traps around the $\hat{\mb{y}}$ axis, we take $k_x = k_z
\equiv k_\text{trap}$ and $k_y = \alpha k_\text{trap}$, where the
weaker axial trapping is reduced by a factor $\alpha <
1$~\cite{Neuman04}.  We assume both traps have equal strengths, though
our method can be generalized to an asymmetric configuration, 
  where the two traps have different strengths $k_\text{trap,1} \ne
  k_\text{trap,2}$.  In this case the reconstruction procedure derived
  below is valid with the substitution $k_\text{trap} = 2
  k_\text{trap,1} k_\text{trap,2}/(k_\text{trap,1}+k_\text{trap,2})$.

We rewrite the Hamiltonian in Eq.~\eqref{eq1} by
defining a total end-to-end coordinate $\mb{r}_\text{tot} \equiv
\mb{r}_2 - \mb{r}_1 = (x_\text{tot},y_\text{tot},z_\text{tot})$, and a
total center-of-mass coordinate $\mb{R}_\text{tot} \equiv \mb{r}_2 +
\mb{r}_1 = (X_\text{tot},Y_\text{tot},Z_\text{tot})$.  In terms of
these variables, ${\cal H}_\text{trap}$ becomes:
\begin{equation}\label{eq2}
\begin{split}
{\cal H}_\text{trap}(\mb{r}_1,\mb{r}_2) &= {\cal H}^\text{cm}_\text{trap}(\mb{R}_\text{tot}) + {\cal H}^\text{ee}_\text{trap}(\mb{r}_\text{tot}),\\
{\cal H}^\text{cm}_\text{trap}(\mb{R}_\text{tot}) &= \frac{1}{4}k_x X_\text{tot}^2 + \frac{1}{4}k_y Y_\text{tot}^2 + \frac{1}{4}k_z (z_\text{trap}-Z_\text{tot})^2,\\
{\cal H}^\text{ee}_\text{trap}(\mb{r}_\text{tot}) &= \frac{1}{4}k_x x_\text{tot}^2 + \frac{1}{4}k_y y_\text{tot}^2 + \frac{1}{4}k_z (z_\text{trap}-z_\text{tot})^2.
\end{split}
\end{equation}
The variables $z_\text{tot}$ and $z_\text{trap}$ are explicitly
labeled in Fig.~1 of the main text.

\subsection{Equilibrium distribution of the system} The equilibrium
probability ${\cal P}_\text{tot}(\mb{R}_\text{tot},\mb{r}_\text{tot})$
of finding the beads at positions with a given $\mb{R}_\text{tot}$ and
$\mb{r}_\text{tot}$ can be expressed as:
\begin{equation}\label{eq3}
{\cal P}_\text{tot}(\mb{R}_\text{tot},\mb{r}_\text{tot}) = A e^{-\beta {\cal H}^\text{cm}_\text{trap}(\mb{R}_\text{tot}) -\beta {\cal H}^\text{ee}_\text{trap}(\mb{r}_\text{tot})} {\cal Q}_\text{tot}(\mb{r}_\text{tot}),
\end{equation}
where $\beta = 1/k_B T$, $A$ is a normalization constant, and ${\cal
  Q}_\text{tot}(\mb{r}_\text{tot})$ is the equilibrium probability of
the total bead-handle-protein system having bead separation
$\mb{r}_\text{tot}$ in the absence of the external trapping potential
or any applied force.  By translational symmetry ${\cal Q}_\text{tot}$
is independent of the center-of-mass coordinates, and by rotational
symmetry ${\cal Q}_\text{tot}(\mb{r}_\text{tot}) = {\cal
  Q}_\text{tot}(| \mb{r}_\text{tot}|)$.  Thus, if we introduce
cylindrical coordinates $\mb{r}_\text{tot} = (\rho_\text{tot},
\phi_\text{tot}, z_\text{tot})$, where $\rho_\text{tot} =
\sqrt{x_\text{tot}^2+y_\text{tot}^2}$, $\phi_\text{tot} =
\tan^{-1}(y_\text{tot}/ x_\text{tot})$, there is no angular
dependence, so that ${\cal Q}_\text{tot}(\mb{r}_\text{tot}) = {\cal
  Q}_\text{tot}(\rho_\text{tot}, z_\text{tot})$.  We are ultimately
interested in the marginal probability ${\cal
  P}_\text{tot}(z_\text{tot})$, which can be derived from the
experimental time series and forms the starting point of our
theoretical procedure to obtain the desired free energy profile.
We obtain ${\cal P}_\text{tot}(z_\text{tot})$ from ${\cal
  P}_\text{tot}(\mb{R}_\text{tot},\mb{r}_\text{tot})$ by integrating
over the $\mb{R}_\text{tot}$, $\rho_\text{tot}$ and $\phi_\text{tot}$
degrees of freedom:
\begin{equation}\label{eq4}
\begin{split}
{\cal P}_\text{tot}(z_\text{tot}) &\equiv \int \rho_\text{tot}\, d\rho_\text{tot}\, d\phi_\text{tot}\,d\mb{R}_\text{tot}\, {\cal P}_\text{tot}(\mb{R}_\text{tot},\mb{r}_\text{tot})\\
&= B \int \rho_\text{tot}\, d\rho_\text{tot}\, d\phi_\text{tot}\, e^{-\beta {\cal H}^\text{ee}_\text{trap}(\rho_\text{tot}, \phi_\text{tot},  z_\text{tot})}{\cal Q}_\text{tot}(\rho_\text{tot},z_\text{tot})\\
&= 2\pi B \int \rho_\text{tot}\, d\rho_\text{tot}\, e^{-\frac{1}{8}\beta(k_x + k_y)\rho_\text{tot}^2 - \frac{1}{4}\beta k_z (z_\text{trap}-z_\text{tot})^2} I_0\left(\frac{1}{8}(k_x-k_y)\rho_\text{tot}^2\right)  {\cal Q}_\text{tot}(\rho_\text{tot},z_\text{tot})\\
&\approx C e^{- \frac{1}{4}\beta k_z (z_\text{trap}-z_\text{tot})^2} {\cal Q}_\text{tot}(z_\text{tot}).
\end{split}
\end{equation}
Here $B$ and $C$ are constants that have absorbed the result of
integrating over $\mb{R}_\text{tot}$ and $\rho_\text{tot}$
respectively, and $I_0$ is a modified Bessel function of the first
kind.  Up to the third line the calculation in Eq.~\eqref{eq4} is
exact.  In the last step we make the problem fully one-dimensional, by
approximately relating ${\cal P}_\text{tot}(z_\text{tot})$ to ${\cal
  Q}_\text{tot}(z_\text{tot})$, defined as ${\cal
  Q}_\text{tot}(z_\text{tot}) = \int \rho_\text{tot}
d\rho_\text{tot}\,{\cal Q}_\text{tot}(\rho_\text{tot},z_\text{tot})$.
We are forced to make this crucial approximation, because
experiments have access only to the $\hat{\mb{z}}$ fluctuations
through ${\cal P}_\text{tot}(z_\text{tot})$, but generally do not have
complete information about the transverse components.  As mentioned in
the main text, the last step in Eq.~\eqref{eq4} becomes exact if: (i)
$k_x = k_y = 0$; or (ii) when ${\cal
  Q}_\text{tot}(\rho_\text{tot},z_\text{tot})$ is separable in the
form ${\cal Q}_\text{tot}(\rho_\text{tot},z_\text{tot}) =
f(\rho_\text{tot}){\cal Q}_\text{tot}(z_\text{tot})$ for some function
$f$.  Though condition (ii) is not expected to be generally valid, we
can approximately satisfy (i) when $\beta
k_\text{trap}\bar{\rho}_\text{tot}^2 \ll 1$, where
$\bar\rho_\text{tot}$ is the typical length scale of total system
fluctuations transverse to $\hat{\mb{z}}$.  Thus, for sufficiently
soft traps, we have in Eq.~\eqref{eq4} a useful relation between the
$\hat{\mb{z}}$ marginal probabilities of the total system with and
without the external trapping potentials.

\subsection{Convolution} Since ${\cal Q}_\text{tot}(z_\text{tot})$ is the
total end-to-end $z$-component distribution in the absence of any
external trapping potential or applied force, the corresponding
distribution for the total system with constant tension $F_0$ applied
to the beads along $\hat{\mb{z}}$ is given by $\tilde{\cal
  P}_\text{tot}(z_\text{tot};F_0) = \exp(\beta F_0 z_\text{tot}) {\cal
  Q}_\text{tot}(z_\text{tot})$.  Substituting for ${\cal
  Q}_\text{tot}(z_\text{tot})$ using Eq.~\eqref{eq4}, we find the
following relation for $\tilde{\cal P}_\text{tot}(z_\text{tot};F_0)$,
which constitutes Step 1 of our construction procedure in the main
text:
\begin{equation}\label{eq4b}
\tilde{\cal P}_\text{tot}(z_\text{tot};F_0) \approx C^{-1} e^{\beta F_0 z + \frac{1}{4}\beta k_z
(z_\text{trap}-z_\text{tot})^2} {\cal P}_\text{tot}(z_\text{tot}).
\end{equation}
The quantity ${\cal P}_\text{tot}(z_\text{tot})$ on the right-hand
side can be derived from the experimental time series, and thus
Eq.~\eqref{eq4b} allows us to obtain an equilibrium distribution in
the constant force ensemble, $\tilde{\cal
  P}_\text{tot}(z_\text{tot};F_0)$, directly from the folding
trajectories.

In the constant force ensemble, $\tilde{\cal P}_\text{tot}$ is just a
1D convolution of the probabilities of the individual system components:
\begin{equation}\label{eq5}
\tilde{\cal P}_\text{tot} = \tilde{\cal P}_\text{b} \ast \tilde{\cal P}_\text{h} \ast \tilde{\cal P}_\text{p} \ast \tilde{\cal P}_\text{h} \ast \tilde{\cal P}_\text{b},
\end{equation}
where $\ast$ denotes the convolution operator.  The probability
$\tilde{\cal P}_{\lambda}(z_{\lambda};F_0)$ is the equilibrium
distribution of $z_{\lambda}$ at constant force $F_0$, where $\lambda$
denotes a bead, handle, or protein.  The quantity $z_{\lambda}$ is the
end-to-end distance of $\lambda$ along $\hat{\mb{z}}$.  Using the
notation in Fig.~1 of the main text, we can give a few examples: for
the protein $z_\text{p} = (\mb{r}_{p2} - \mb{r}_{p1})\cdot
\hat{\mb{z}}$; for the left handle $z_\text{h} = (\mb{r}_{p1} -
\mb{r}_{1}^\prime)\cdot \hat{\mb{z}}$; for the left bead $z_\text{b} =
(\mb{r}_1^\prime - \mb{r}_1)\cdot \hat{\mb{z}}$.  In Fourier space
Eq.~\eqref{eq5}, which is the key equation for Step 2 of the
construction procedure in the main text, has a simple form:
\begin{equation}\label{eq6}
\tilde{\cal P}_\text{tot}(k;F_0) = \tilde{\cal P}_\text{b}^2(k;F_0) \tilde{\cal P}^2_\text{h}(k;F_0) \tilde{\cal P}_\text{p}(k;F_0) \equiv \tilde{\cal P}_\text{bh}(k;F_0) \tilde{\cal P}_\text{p}(k;F_0),
\end{equation}
where $\tilde{\cal P}_\lambda(k;F_0)$ is the Fourier transform of
$\tilde{\cal P}_\lambda(z_\lambda;F_0)$.  Here $\tilde{\cal
    P}_\text{bh}(k;F_0) = \tilde{\cal P}_\text{b}^2(k;F_0) \tilde{\cal
    P}^2_\text{h}(k;F_0)$ is the Fourier transform of the convolution
  of all the bead and handle distributions. If the left and right
  handles (or analogously the beads) had distinct properties
  (i.e. different sizes) then the factor $\tilde{\cal
    P}^2_\text{h}(k;F_0)$ in $\tilde{\cal P}_\text{bh}(k;F_0)$ would
  be replaced by the product $\tilde{\cal
    P}_\text{h1}(k;F_0)\tilde{\cal P}_\text{h2}(k;F_0)$ of the
  distinct handle terms.  Given the rotational properties of the
beads and modeling the handles as semiflexible polymers, we can derive
a numerically exact form for the Fourier components $\tilde{\cal
  P}_\text{bh}(k;F_0)$, and hence by inversion the corresponding real
space distribution.  This will allow us to directly recover
$\tilde{\cal P}_\text{p}$ from $\tilde{\cal P}_\text{tot}$, without
resorting to an experimental estimate for the point spread function,
which is problematic due to the varying force conditions that arise in
optical traps with non-zero stiffness.

\subsection{Bead distribution} \label{sec:bead} The first
step in finding $\tilde{\cal P}_\text{bh}(k;F_0) = \tilde{\cal
  P}_\text{b}^2(k;F_0) \tilde{\cal P}^2_\text{h}(k;F_0)$ is to obtain an
expression for the Fourier-space bead probability $\tilde{\cal
  P}_\text{b}(k;F_0)$.  Taking as an example the left bead in Fig.~1,
let $\mb{r}_\text{b} = \mb{r}_1^\prime - \mb{r}_1$ be the vector
between the bead center and the point on the bead surface that is
attached to the handle.  This vector has a fixed length $R_b$ given by
the bead radius, but its direction can fluctuate, subject to a
constant force $F_0$ along $\hat{\mb{z}}$.  The equilibrium
distribution $\tilde{\cal P}_\text{b}(\mb{r}_\text{b};F_0)$ is given
by:
\begin{equation}\label{eq7}
\tilde{\cal P}_\text{b}(\mb{r}_\text{b};F_0) =
A_\text{b} e^{\beta F_0 z_\text{b}} \delta\left(|\mb{r}_\text{b}| - R_b \right),
\end{equation}
with the delta function enforcing the constraint $|\mb{r}_\text{b}|
= R_b$, and the normalization constant $A_\text{b}$.  The quantity
$\tilde{\cal P}_\text{b}(k;F_0)$ is the Fourier transform of
$\tilde{\cal P}_\text{b}(\mb{r}_\text{b};F_0)$ evaluated at
$\mb{k} = k \hat{\mb{z}}$:
\begin{equation}\label{eq8}
\begin{split}
\tilde{\cal P}_\text{b}(k;F_0) &= \int d\mb{r}_\text{b}\,e^{-i k z_\text{b}} \tilde{\cal P}_\text{b}(\mb{r}_\text{b};F_0)\\
&= \frac{\beta F_0 \sinh\left((\beta F_0-ik)R_b\right)}{(\beta F_0-ik) \sinh\left(\beta F_0 R_b\right)}.
\end{split}
\end{equation}

\subsection{Handle distribution} \label{sec:hand} Though the
Fourier components $\tilde{\cal P}_\text{h}(k;F_0)$ of the semiflexible
handle distribution do not have a simple analytic expression, they can
be calculated numerically to arbitrary accuracy.  The Hamiltonian for
the semiflexible handle polymer with contour length $L$, persistence
length $l_p$, and elastic stretching modulus $\gamma$, can be exactly
mapped onto the propagator of a quantum particle on the
surface of a unit sphere~\cite{Samuel02,Kierfeld04}.  Following the
approach in Ref.~\cite{Kierfeld04}, we describe the polymer as a
continuous spatial contour $\mb{r}(s)$ in terms of an unstretched arc
length $s$ which runs from $s=0$ to $s=L$. At each point $s$ we define
a unit tangent vector $\mb{u}(s)$.  The end-to-end distance $\mb{r}(L)
- \mb{r}(0)$ can be written as,
\begin{equation}\label{qu0}
\mb{r}(L) - \mb{r}(0) = \int_0^L ds\, (1+\epsilon(s)) \mb{u}(s),
\end{equation}
where $1+\epsilon(s)$ is the local relative bond length extension.
For an inextensible ($\gamma \to \infty$) worm-like chain,
$\epsilon(s) = 0$ for all $s$, which corresponds to all bonds in the
chain having fixed length.  For finite $\gamma$, the $\epsilon(s)$ are
additional degrees of freedom in the system, which together with the
unit tangent vectors $\mb{u}(s)$ completely define the contour.  The
Hamiltonian ${\cal H}(\mb{u}(s),\epsilon(s))$ for the semiflexible
polymer under tension is,
\begin{equation}\label{qu1}
\begin{split}
\beta {\cal H}(\mb{u}(s),\epsilon(s)) &= \frac{l_p}{2} \int_0^L ds\, (\partial_s \mb{u}(s))^2 - f\hat{\mb{z}}\cdot (\mb{r}(L)-\mb{r}(0)) +  \frac{\beta\gamma}{2} \int_0^L ds\, \epsilon^2(s),\\
&= \int_0^L ds\,\left[ \frac{l_p}{2} (\partial_s \mb{u}(s))^2 - f(1+\epsilon(s))u_z(s) +  \frac{\beta\gamma}{2} \epsilon^2(s)\right],
\end{split}
\end{equation}
where $\beta = 1/k_B T$ and we have used Eq.~\eqref{qu0} for the
end-to-end distance.  The first term in Eq.~\eqref{qu1} corresponds to
a bending energy parameterized by the persistence length $l_p$, the
second term is due to an applied mechanical force $k_B T f$ along
$\hat{\mb{z}}$, and the third term describes the stretching energy of
the bonds, with elastic modulus $\gamma$.  For prestretching tension
$F_0$, $f=\beta F_0$, but for convenience we will extend the
definition of ${\cal H}$ to include arbitrary $f$ in order to obtain
the Fourier components of the end-to-end probability distribution
below.

The partition function of the polymer (with free end boundary
conditions) can be expressed as a path integral over all possible
configurations of $\mb{u}(s)$ and $\epsilon(s)$, with the constraint
that $\mb{u}^2(s) = 1$ at each $s$:
\begin{equation}\label{qu1b}
\begin{split}
Z_\text{h}(f) &= \int {\cal D}\mb{u}(s) \prod_s \delta(\mb{u}^2(s)-1)\, \int {\cal D}\mb{\epsilon}(s)\,\exp\left[-\beta {\cal H}(\mb{u}(s),\epsilon(s))\right],\\
&\equiv \int {\cal D}\mb{u}(s) \prod_s \delta(\mb{u}^2(s)-1)\, \exp\left[-\beta {\cal H}_\text{eff}(\mb{u}(s))\right]
\end{split}
\end{equation}
up to some normalization constant.  In the second line we have 
carried out the path integral over $\epsilon(s)$ exactly to express
$Z_\text{h}(f)$ in terms of an effective Hamiltonian ${\cal
  H}_\text{eff}(\mb{u}(s))$ depending on the tangent vectors alone,
\begin{equation}\label{qu1c}
\beta {\cal H}_\text{eff}(\mb{u}(s)) = \int_0^L ds\,\left[ \frac{l_p}{2} (\partial_s \mb{u}(s))^2 - f u_z(s) -  \frac{f^2}{2\beta\gamma} u_z^2(s)\right].
\end{equation}
The probability of finding the polymer in a configuration with an
end-to-end extension $z_\text{h}$ along
$\hat{\mb{z}}$ is given by~\cite{Samuel02}:
\begin{equation}\label{qu1d}
\begin{split}
\tilde{\cal P}_\text{h}(z_\text{h};F_0) &= \frac{1}{Z_\text{h}(\beta F_0)} \int {\cal D}\mb{u}(s) \prod_s \delta(\mb{u}^2(s)-1)\,\\
&\qquad \int {\cal D}\mb{\epsilon}(s)\,\delta\left( z_{\text{hand}} - \int_0^L ds\,(1+\epsilon(s))u_z(s) \right) \exp\left[-\beta {\cal H}(\mb{u}(s),\epsilon(s))\right]\\
&= \frac{1}{Z_\text{h}(\beta F_0)} \int {\cal D}\mb{u}(s) \prod_s \delta(\mb{u}^2(s)-1)\,\\
&\qquad \int {\cal D}\mb{\epsilon}(s)\,\int \frac{dk}{2\pi} e^{ik\left( z_\text{h} - \int_0^L ds\,(1+\epsilon(s))u_z(s) \right)} \exp\left[-\beta {\cal H}(\mb{u}(s),\epsilon(s))\right]\\
&\equiv \int \frac{dk}{2\pi}\, e^{ik z_\text{h}} \tilde{\cal P}_\text{h}(k;F_0),
\end{split}
\end{equation}
where the Fourier components of the probability distribution are:
\begin{equation}\label{eq9}
\tilde{\cal P}_\text{h}(k;F_0) = \frac{Z_\text{h}(\beta F_0 - i k)}{Z_\text{h}(\beta F_0)}.
\end{equation}

In order to evaluate $\tilde{\cal P}_\text{h}(k;F_0)$, we need to
calculate $Z_\text{h}(f)$.  Let us define the propagator
$G(\mb{u}_0, \mb{u}_{L}; L)$ as the path integral over all
configurations with initial tangent $\mb{u}(0) = \mb{u}_0$ and final
tangent $\mb{u}(L) = \mb{u}_{L}$:
\begin{equation}\label{qu3}
G(\mb{u}_0, \mb{u}_{L}; L) = \int_{\mb{u}(0) =
\mb{u}_0}^{\mb{u}(L) =
\mb{u}_{L}} {\cal D}\mb{u}(s) \prod_s
\delta(\mb{u}^2(s)-1)\, e^{-\beta {\cal H}_\text{eff}(\mb{u}(s))}\,.
\end{equation}
This is related to the partition function through $Z_\text{h}(f) = (4\pi)^{-2}
\int_S d\mb{u}_0\, d\mb{u}_{L}\, G(\mb{u}_0, \mb{u}_{L}; L)$, where the integrations are over the unit sphere $S$.

The quantum Hamiltonian corresponding to $\beta {\cal H}_\text{eff}$ is
\begin{equation}\label{qu4}
{\cal H}^\text{qu}_\text{eff}(f) = -\frac{1}{2l_p}\nabla^2 - f \cos \theta - \frac{f^2}{2\beta \gamma} \cos^2\theta\,,
\end{equation}
describing a particle on the surface of a unit sphere, with $\theta
=0$ defining the $\hat{\mb{z}}$ direction.  The propagator $G$ can be
written in terms of the quantum eigenvalues $E_n$ and
eigenstates $\psi_n(\mb{u})$ of ${\cal H}^\text{qu}_\text{eff}$:
\begin{equation}\label{qu5}
G(\mb{u}_0, \mb{u}_{L}; {L}) = \sum_n e^{-E_n {L}} \psi_n^\ast(\mb{u}_0) \psi_n(\mb{u}_{L}) = \sum_{n,l,m,l^\prime,m^\prime} e^{-E_n {L}} a^\ast_{nl^\prime m^\prime}a_{nlm}Y_{l^\prime m^\prime}^\ast(\mb{u}_0) Y_{lm}(\mb{u}_{L})\,,
\end{equation}
where we have expanded the eigenstates in the basis of spherical
harmonics, $\psi_n(\mb{u}) = \sum_{lm} a_{nlm} Y_{lm}(\mb{u})$. The
coefficients $a_{nlm}$ are the components of the $n$th
eigenvector of the Hamiltonian ${\cal H}_\text{eff}^\text{qu}$ in the
$Y_{lm}$ basis.  The partition function $Z_\text{h}(f)$ becomes:
\begin{equation}\label{qu7}
\begin{split}
Z_\text{h}(f) &= \frac{1}{(4\pi)^2} \int_S d\mb{u}_0\, d\mb{u}_{L}\, \sum_{n,l,m,l^\prime,m^\prime} e^{-E_n {L}} a^\ast_{nl^\prime m^\prime}a_{nlm}Y_{l^\prime m^\prime}^\ast(\mb{u}_0) Y_{lm}(\mb{u}_{L})\\
&= \frac{1}{4\pi} \sum_n e^{-E_n {L}} a^\ast_{n00}a_{n00}\\
&= \langle 0 | e^{-L {\cal H}_\text{eff}^\text{qu}(f)} | 0 \rangle.
\end{split}
\end{equation}
In the last step we have written the expression as a single component
of the exponentiated matrix ${\cal H}_\text{eff}^\text{qu}(f)$ in the
  $(l,m)$ spherical harmonic basis, where $|l\rangle$ denotes a state
  $(l, 0)$.  Since the Hamiltonian matrix in the $m=0$ subspace does
  not couple to $m\ne 0$ components, we only need $m=0$ matrix
  elements to evaluate $Z_\text{h}(f)$.  The list of non-zero
  matrix entries in the $m=0$ subspace is:
\begin{equation}\label{eq10}
\begin{split}
\langle l | {\cal H}_\text{eff}^\text{qu}(f) | l \rangle &= \frac{1}{2l_p} l(l+1) - \frac{f^2}{2\beta\gamma} \frac{2 l^2+2 l-1}{(2 l-1) (2 l+3)}, \qquad l=0,1,2,\ldots, \\
\langle l | {\cal H}_\text{eff}^\text{qu}(f) | l+1 \rangle &= \langle l+1 | {\cal H}_\text{eff}^\text{qu}(f) | l \rangle
= \frac{f (l+1)}{\sqrt{(2l+1)(2l+3)}}, \qquad l=0,1,2,\ldots,\\
\langle l | {\cal H}_\text{eff}^\text{qu}(f) | l+2 \rangle &= \langle l+2 | {\cal H}_\text{eff}^\text{qu}(f) | l \rangle
= -\frac{f^2}{2\beta\gamma} \frac{(l+1) (l+2)}{(2 l+3) \sqrt{(2 l+1) (2 l+5)}}, \qquad l=0,1,2,\ldots.\\
\end{split}
\end{equation}
To carry out the matrix exponent, we truncate the matrix at
$l_\text{max}=20$, which is sufficiently large for numerical
accuracy.

\begin{figure}[t]
\centering\includegraphics*[scale=1]{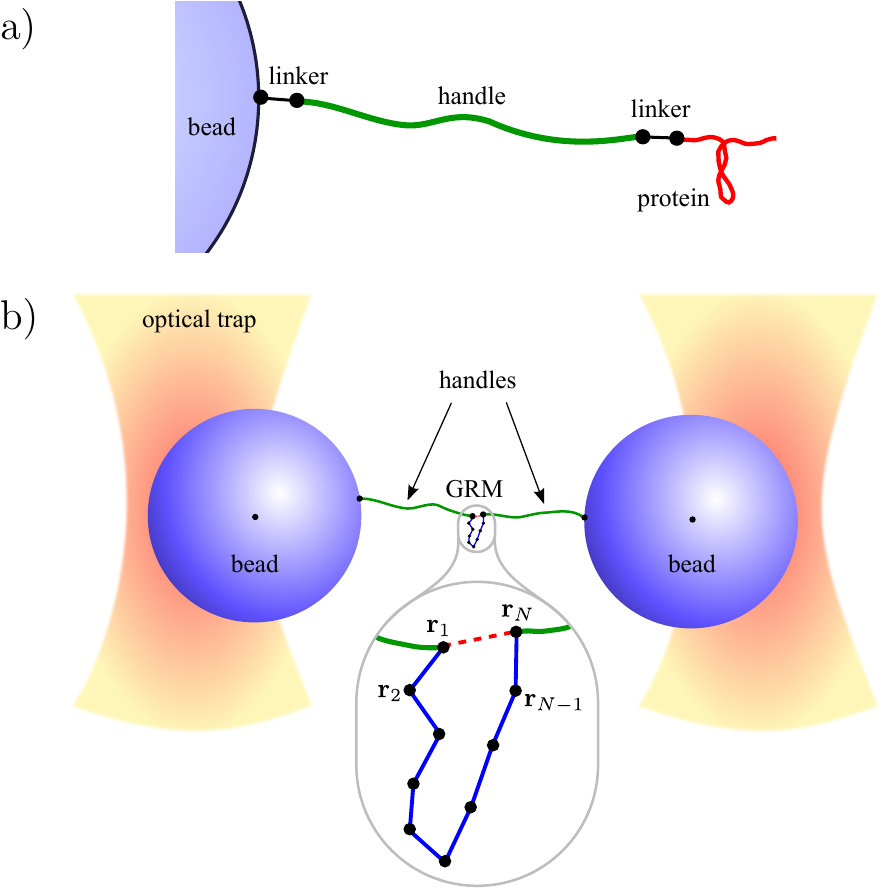}
  \caption{a) Schematic illustration (not to scale) of the covalent
    linkers which attach the handle to the bead on one end, and the
    handle to the protein on the other end. b) The Generalized Rouse
    Model (GRM) within the optical tweezer.}\label{link}
\end{figure}

In some experimental setups, covalent linkers are attached on both
ends of each handle, connecting the handle to the neighboring bead and
protein, as schematically drawn in Fig.~\ref{link}(a).  The effect of
linkers can be absorbed into the theory by modifying
$\tilde{P}_\text{h}(k;F_0)$.  The simplest representation of
a linker is a harmonic spring with stiffness $\kappa$ and
natural length $\ell$.  With one of these added at each end of the
handle, Eq.~\eqref{eq9} becomes:
\begin{equation}\label{eq10b}
\tilde{\cal P}_\text{h}(k;F_0) = \frac{Z_\text{h}(\beta F_0 - i k)}{Z_\text{h}(\beta F_0)} \frac{Z_\text{link}^2(\beta F_0 - i k; \kappa, \ell)}{Z_\text{link}^2(\beta F_0; \kappa, \ell)},
\end{equation}
where
\begin{equation}\label{eq10c}
\begin{split}
Z_\text{link}(f; \kappa, \ell) &= \frac{1}{f}\sqrt{\frac{\pi }{2 (\beta \kappa)^3}} e^{\frac{f (f-2 \ell \beta \kappa )}{2 \beta \kappa }} \Biggl[\ell \beta \kappa 
   \,\text{erf}\left(\frac{f-\ell \beta \kappa }{\sqrt{2 \beta \kappa }}\right)\\
&\qquad +e^{2 f \ell} (f+\ell \beta \kappa )
   \left(\text{erf}\left(\frac{f+\ell \beta \kappa }{\sqrt{2 \beta \kappa }}\right)+1\right)
+f
   \text{erfc}\left(\frac{f-\ell \beta \kappa }{\sqrt{2 \beta \kappa }}\right)-\ell \beta \kappa \Biggr].
\end{split}
\end{equation}

\subsection{Numerical deconvolution to extract the protein distribution}\label{sec:numdec}

The expressions given by Eq.~\eqref{eq8} and \eqref{eq10b} completely
determine the Fourier-transformed point spread function $\tilde{\cal
  P}_\text{bh}(k;F_0)$ at all $k$.  Naively, one could use
Eq.~\eqref{eq6} to write:
\begin{equation}\label{ext1}
\tilde{\cal P}_\text{p}(k;F_0) = \frac{\tilde{\cal
  P}_\text{tot}(k;F_0)}{\tilde{\cal P}_\text{bh}(k;F_0)}.
\end{equation}
Since $\tilde{\cal
  P}_\text{tot}(k;F_0)$ is derivable from the experimental times series,
this would immediately yield $\tilde{\cal P}_\text{p}(k;F_0)$, and
after inversion the ultimate goal, $\tilde{\cal P}_\text{p}(
z_\text{p};F_0)$.  However, this direct deconvolution in Fourier space
is numerically unstable~\cite{Press07}, due to the effects of
round-off noise and the denominator in the equation for $\tilde{\cal
  P}_\text{p}(k;F_0)$ approaching zero at large $k$.

To work around this problem, we implement the deconvolution in real
space, by solving the following integral equation for
$\tilde{\cal P}_\text{p}$ (the real space version of
Eq.~\eqref{eq6}):
\begin{equation}\label{eq11}
\int d z_\text{p}\,\tilde{\cal P}_\text{bh}(z_\text{tot} - z_\text{p};F_0) \tilde{\cal P}_\text{p}(z_\text{p};F_0) = \tilde{\cal P}_\text{tot}(z_\text{tot};F_0).
\end{equation}
One way to approach Eq.~\eqref{eq11} is to approximate the integral as
a matrix-vector product by discretizing the $z_\text{tot}$ and
$z_\text{p}$ ranges.  However, the convolution matrix corresponding to
$\tilde{\cal P}_\text{bh}$ is generally ill-conditioned, so direct
inversion to find a solution is unfeasible.  Alternatively, to obtain
robust, smooth results for the deconvolution, we can rewrite
Eq.~\eqref{eq11} by representing the three quantities $\tilde{\cal
  P}_\text{p}$, $\tilde{\cal P}_\text{bh}$, and $\tilde{\cal
  P}_\text{tot}$ in terms of suitable fitting functions.  Since these
are all probability distributions, in practice we can approximate them
to arbitrary precision as sums of Gaussians $g(z;\zeta,v) = (2\pi
v)^{-1/2} \exp(-(z-\zeta)^2/(2v))$,
\begin{equation}\label{eq12}
 \tilde{\cal
  P}_\alpha(z_\alpha;F_0) = \sum_{i=1}^{N_\alpha} a^\alpha_i  g(z_\alpha;\zeta^\alpha_i,v^\alpha_i),
\end{equation}
where $\alpha = \text{p}$, bh, or tot.  The number of Gaussians needed
for each distribution, $N_\alpha$, is chosen depending on the problem.
The two sets of parameters
$\{a^\text{tot}_i,\xi^\text{tot}_i,v^\text{tot}_i\}$ and
$\{a^\text{bh}_i,\xi^\text{bh}_i,v^\text{bh}_i\}$ (which implicitly
depend on $F_0$) are computed by fitting to the known functions
$\tilde{\cal P}_\text{tot}$ and $\tilde{\cal P}_\text{bh}$.  The goal
of the procedure is then to use Eq.~\eqref{eq11} to solve for the
parameter set $\{a^\text{p}_i,\xi^\text{p}_i,v^\text{p}_i\}$
describing the unknown function $\tilde{\cal P}_\text{p}$.  For the
cases discussed in the main text, choosing $N_\alpha=2-3$ was
sufficient to find solutions $\tilde{\cal P}_\text{p}$ such that the
left and right-hand sides of Eq.~\eqref{eq11} had a median deviation
$\lesssim 1\%$ over all $z_\text{tot}$ where ${\cal
  P}_\text{tot}(z_\text{tot}) \gtrsim 10^{-6}$.

The details of the solution procedure are as follows: we choose
$N_\text{p} = N_\text{tot}$, so that Eq.~\eqref{eq11} can be
approximated as a one-to-one convolution mapping each Gaussian in
$\tilde{\cal P}_\text{p}$ into a corresponding Gaussian in
$\tilde{\cal P}_\text{tot}$.  For all $i = 1,\ldots,N_\text{tot}$,
Eq.~\eqref{eq11} describes the following relationships between the
amplitudes, positions and variances of the Gaussians:
\begin{equation}\label{eq13}
\begin{split}
a^\text{tot}_i &\approx a^\text{p}_i,\\
\xi^\text{tot}_i &\approx \sum_{j=1}^{N_\text{bh}} a^\text{bh}_j (\xi^\text{bh}_j+ \xi^\text{p}_i),\\
v^\text{tot}_i &\approx \sum_{j=1}^{N_\text{bh}} a^\text{bh}_j (v^\text{bh}_j+ v^\text{p}_i + (\xi^\text{bh}_j+ \xi^\text{p}_i)^2).\\
\end{split}
\end{equation}
The approximation is exact when the point-spread function $\tilde{\cal
  P}_\text{bh}$ is precisely a single Gaussian, but is generally valid
whenever $\tilde{\cal P}_\text{bh}$ is close to Gaussian (as is the
case for the bead-handle system, where the corrections introduced by
choosing $N_\text{bh}>1$ are small).  Eq.~\eqref{eq13} can be inverted
to yield the desired parameter set
$\{a^\text{p}_i,\xi^\text{p}_i,v^\text{p}_i\}$:
\begin{equation}\label{eq14}
\begin{split}
a^\text{p}_i &\approx a^\text{tot}_i,\\
\xi^\text{p}_i &\approx \xi^\text{tot}_i - \sum_{j=1}^{N_\text{bh}} a^\text{bh}_j \xi^\text{bh}_j,\\
v^\text{p}_i &\approx v^\text{tot}_i - \sum_{j=1}^{N_\text{bh}} a^\text{bh}_j \left[ v^\text{bh}_j + \left(\xi^\text{bh}_j+ \xi^\text{tot}_i - \sum_{k=1}^{N_\text{bh}} a^\text{bh}_k \xi^\text{bh}_k\right)^2\right],\\
\end{split}
\end{equation}
where we have used the fact that $\sum_i^{N_\text{bh}} a^\text{bh}_i
=1$ due to normalization.

\section{Experimental verification of the model for the point-spread function}\label{sec:ps}

\begin{figure}[t]
\centering\includegraphics*[width=0.98\textwidth]{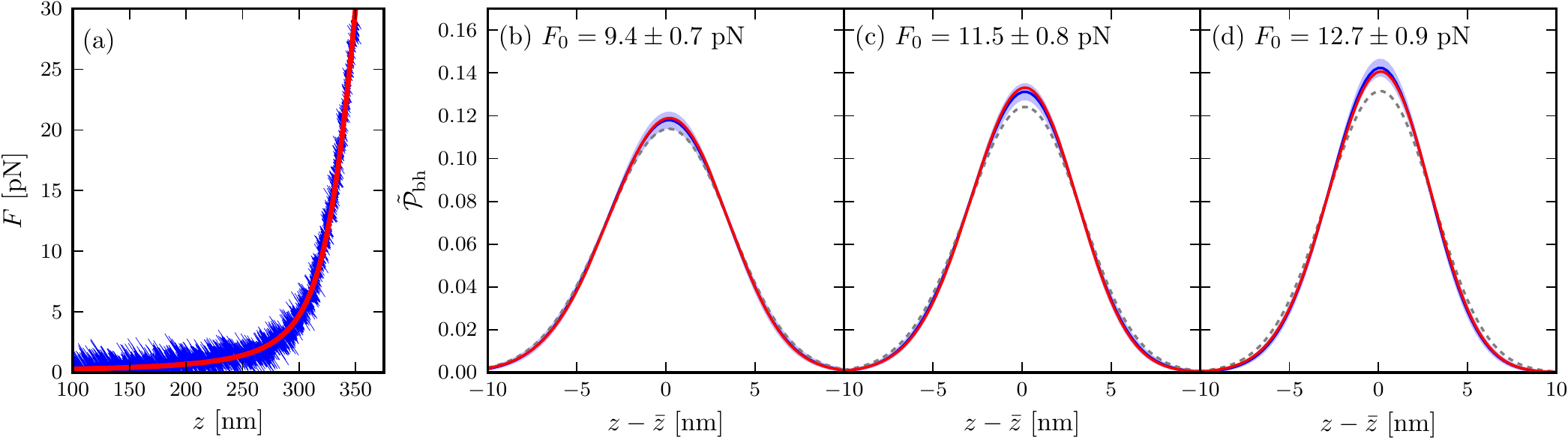}
\caption{Experimental (blue solid curves) and theory (red solid
  curves) results for a system containing only dsDNA handles and
  beads.  The apparatus parameters and theoretical best-fit values are
  described in Sec.~\ref{sec:ps}.  The same set of best-fit parameters
  is used for all the theory curves.  (a) Force $F$ vs. total
  extension $z$ from an experimental pulling trajectory, compared to
  the theoretical mean extension as a function of force; (b-d) Total
  bead-handle probability end-to-end distance distributions
  $\tilde{\cal P}_\text{bh}$ (blue solid curves) derived from three
  experimental runs at different constant trap separations,
  corresponding to mean forces of $9.4\pm 0.7$, $11.5\pm 0.8$, and
  $12.7\pm 0.9$ pN respectively.  In each case the experimental data
  is corrected for noise/filtering effects using the FBS method
  (Sec.~\ref{sec:fbs}), and transformed into the constant force
  ensemble using Eq.~[1] of the main text, with $F_0$ chosen to be
  equal to the mean force value in the trajectory.  The distance scale
  is centered at $\bar{z}$, the mean extension.  The light blue shaded
  region around each blue curve represents the standard error margin
  for every point in the distribution (68\% confidence band).  For
  comparison, the experimental results omitting the FBS corrections
  are shown as gray dashed lines.}\label{hand_test}
\end{figure}

In order to check that the theoretical model of the point-spread
function $\tilde{\cal P}_\text{bh}$ derived in
Secs.~\ref{sec:bead}-\ref{sec:hand} is an accurate description of the
handle and bead response in experiments, we analyzed control
experiments on a system with only dsDNA handles and beads.  Bead radii
are $R_b = 500 \pm 25$ nm, the trap strength is $k_\text{trap} = 0.29
\pm 0.02$ pN/nm, and the handle parameters are extracted from the
theoretical best-fit described below.  Four distinct experimental data
sets are collected (Fig.~\ref{hand_test}): the first is from a pulling
setup, where the trap separation is varied to give a trajectory of
force $F$ vs. total extension $z$ (Fig.~\ref{hand_test}(a), blue
curve); the other data sets are trajectories of extension $z$ as a
function of time collected at three different constant trap
separations.  These three trajectories can be binned, and projected
onto the constant force ensemble using the same method
(Eq.~\eqref{eq4b}) as described above for the full system, yielding
probability distributions $\tilde{\cal P}_\text{bh}(z;F_0)$ for the total
end-to-end extension of the bead-handle system
(Fig.~\ref{hand_test}(b-d), blue curves).  The constant force value
$F_0$ for each projection is chosen equal to the mean force in each of
the three trajectories, namely $F_0 = 9.4\pm 0.7$, $11.5\pm 0.8$, and
$12.7\pm 0.9$ pN.

The experimental data were collected at 100 kHz, with no additional
time averaging beyond the electronic filtering intrinsic to the
detection and recording apparatus.  Prior to the projection onto the
constant force ensemble, the FBS method (Sec.~\ref{sec:fbs}) was used
to approximately correct the raw experimental data for distortions due
to electronic filtering and noise.  In the absence of these
corrections, the $\tilde{\cal P}_\text{bh}(z;F_0)$ from the raw data is
given by the dashed curves in Fig.~\ref{hand_test}(b-d).

The standard error margin (68\% confidence interval) for each point in
the $\tilde{\cal P}_\text{bh}$ distribution is marked by a light blue
band, reflecting uncertainties in apparatus and FBS parameters, as
well as statistical error due to sampling.  Details of the error
estimation procedure are in Sec.~\ref{sec:err}.  The median standard error
in the $z$ range shown varies from $3-5\%$ between the three
trajectories.

We use the theoretical model of Secs.~\ref{sec:bead}-\ref{sec:hand} to
simultaneously fit all four experimental data sets with a single set
of handle parameters, yielding best fit values: $L=173\pm 2$ nm, $l_p
= 11\pm 1$ nm, and $\gamma = 520 \pm 70$ pN.  The theory has excellent
agreement with all the experimental results, with median deviations in
$\tilde{\cal P}_\text{bh}(z;F_0)$ for the $z$ range shown in
Fig.~\ref{hand_test}(b-d) varying from $1-3\%$ between the three
trajectories, comparable to the standard error margins.  The
comparison between theory and experiments firmly establishes the
remarkable accuracy of our theory in quantitatively describing the
bead-handle system.

\section{Generalized Rouse Model (GRM)}

\subsection{Hamiltonian and exact probability distribution for the GRM}\label{sec:grm}

The GRM model~\cite{Hyeon08}, illustrated schematically in
Fig.~\ref{link}(b), is a Gaussian chain with $N$ monomers, connected
by $N-1$ harmonic springs with an average extension $a$.  A
conformation of the GRM is specified by the monomer positions
$\mb{r}_i$, $i=1,\ldots,N$.  To get behavior reminiscent of hairpin
unzipping, an additional harmonic bond potential,
$V(|\mb{r}_N-\mb{r}_1|)$, is added between the end-points $\mb{r}_1$
and $\mb{r}_N$; the force due to this potential is non-zero only if
the end-point separation is within a cutoff distance, $c$.  Under a
constant external tension, $F_0 \hat{\mb{z}}$, the GRM Hamiltonian is
\begin{equation}\label{grmh}
{\cal H}_\text{GRM} = \frac{3k_B T}{2a^2}\sum_{i=1}^{N-1}
(\mb{r}_{i+1}-\mb{r}_i)^2 + V(|\mb{r}_N - \mb{r}_1|) - F_0
\hat{\mb{z}}\cdot (\mb{r}_N - \mb{r}_1),
\end{equation}
where $V(r) = k r^2 \Theta(c-r) + kc^2 \Theta(r-c)$, and $\Theta$ is
the unit step function.  We choose parameters: $N=18$, $F_0 = 2.9$
$k_BT/$nm (11.9 pN), $a = 1$ nm, $c=12$ nm, $k = 0.09$ $k_BT/$nm$^2$
(0.37 pN/nm).  

If we write the end-to-end vector $\mb{r}_N - \mb{r}_1$ in cylindrical
coordinates as $(\rho,\phi,z)$, the exact probability distribution for
this vector in equilibrium under constant force $F_0\hat{\mathbf{z}}$
is given by:
\begin{equation}\label{eq14}
\tilde{\cal P}_\text{GRM}(\rho,\phi,z;F_0) = A_\text{GRM} \exp\left(-\frac{3(\rho^2+z^2)}{2a^2(N-1)} - \beta V(\sqrt{\rho^2+z^2}) + \beta F_0 z\right),
\end{equation}
where $A_\text{GRM}$ is a normalization constant.  This distribution,
projected onto the $(\rho,z)$ plane, is illustrated in the top panel
of Fig.~2(a) in the main text.  The peak at small $z$ corresponds to
the ``folded'' hairpin state (F) with an intact end-point bond, while
the peak at larger $z$ is the unfolded (U) state.  Integrating
$\tilde{\cal P}_\text{GRM}(\rho,\phi,z;F_0)$ over $\rho$ and $\phi$
one obtains the marginal probability $\tilde{\cal
  P}_\text{GRM}(z;F_0)$,
\begin{equation}\label{eq15}
\begin{split}
\tilde{\cal P}_\text{GRM}(z;F_0) &= A^\prime_\text{GRM}
e^{-\frac{3 z^2}{2a^2(N-1)} + \beta F_0 z}\\ \qquad
&\cdot \begin{cases}
  e^{-\frac{c^2(3+2a^2\beta k(N-1))-3z^2}{2a^2(N-1)}} -
  \frac{3e^{-\beta kz^2}}{3+2a^2 \beta k(N-1)}
  \left(e^{-(c^2-z^2)\left(\beta k+\frac{3}{2a^2(N-1)}\right)}-1\right) & z
  \le c\\
e^{-\beta kc^2} & z > c
 \end{cases},
\end{split}
\end{equation}
with normalization constant $A^\prime_\text{GRM}$.  $\tilde{\cal
  P}_\text{GRM}(z;F_0)$ is plotted in the lower panel of Fig.~2(a).

\subsection{Testing the GRM deconvolution at various forces and trap strengths}

\begin{figure}[t]
\centering\includegraphics*[scale=1]{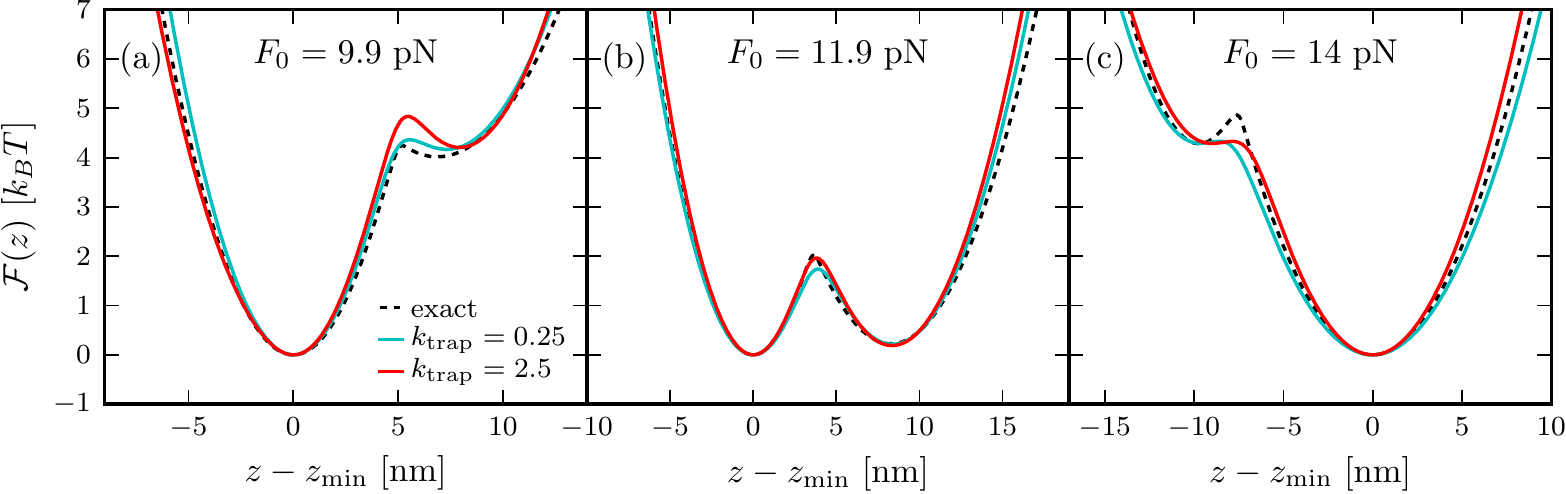}
\caption{The GRM free energy $\tilde{F}_\text{GRM}$ after
  deconvolution for three different values of the force $F_0$: (a) 9.9
  pN; (b) 11.9 pN; (c) 14 pN.  In each case, results for two different
  trap strengths $k_\text{trap} = 0.25$, $2.5$ pN/nm are shown as
  solid lines of different color.  The exact analytical solutions are
  drawn as dashed lines.  The $z$ scale is plotted relative to
  $z_\text{min}$, the location of the minimum in the free energy.}\label{grmextra}
\end{figure}

In Fig.~3(b) in the main text we showed that the deconvolution results
for the GRM are robust when varying the handle parameters.  In
Fig.~\ref{grmextra} we demonstrate that the same conclusion holds when
either the force $F_0$ or the trap strength $k_\text{trap}$ are
varied.

\section{WHAM:  combining trajectories from experimental runs at different trap separations}

The weighted histogram analysis method~\cite{Ferrenberg89} (WHAM) is a
powerful tool in analyzing optical tweezer experiments.  By combining
trajectories generated at different trap separations $z_\text{trap}$
(resulting in different force scales $\bar{F}$), one can sample the
full extent of the protein free energy landscape, and use WHAM to
construct a single energy profile using all the trajectory data, as
has been previously done in Ref.~\cite{Shirts08} (and in a related, but
different manner in Ref.~\cite{Messieres11}.)  In the context of our
theory, WHAM modifies Step 1 of our procedure, allowing us to derive
the equilibrium probability $\tilde{\cal
  P}_\text{tot}(z_\text{tot};F_0)$ at constant force $F_0$ based on
information from multiple experimental trajectories.  Consider a set
of $M$ experimental runs, where the $i$th trajectory consists of $n_i$
data points and has a trap separation $z^{(i)}_\text{trap}$.  Except
for $z^{(i)}_\text{trap}$, all other system parameters are kept the
same between runs.  For each run one can calculate the normalized
histogram of total end-to-end distances $z_\text{tot}$, yielding a
probability distribution ${\cal P}_\text{tot}^{(i)}(z_\text{tot})$.
This distribution is related to ${\cal Q}_\text{tot}(z_\text{tot})$,
the unbiased $z_\text{tot}$ probability in the absence of a trapping
potential or external force, through Eq.~\eqref{eq4}.  Inverting that
equation, we can write
\begin{equation}\label{eqw1}
\begin{split}
 {\cal Q}_\text{tot}(z_\text{tot}) &\approx  C_i^{-1} e^{\frac{1}{4}\beta k_z (z_\text{trap}^{(i)}-z_\text{tot})^2} {\cal P}_\text{tot}^{(i)}(z_\text{tot})\\
&\equiv e^{\beta(U_i(z_\text{tot})-F_i)}{\cal P}_\text{tot}^{(i)}(z_\text{tot}),
\end{split}
\end{equation}
where $C_i$ is a normalization constant, $U_i(z_\text{tot}) =
k_z(z_\text{trap}^{(i)} - z_\text{tot})^2/4$ and $F_i = \beta^{-1} \ln
C_i$.  In the case of one trajectory ($M=1$), Eq.~\eqref{eqw1} is a
way to estimate ${\cal Q}_\text{tot}(z_\text{tot})$, from which one
can calculate $\tilde{\cal P}_\text{tot}(z_\text{tot};F_0) = \exp(\beta
F_0 z_\text{tot}) {\cal Q}_\text{tot}(z_\text{tot})$.  This is just
the standard Step 1 procedure described earlier.

When $M>1$, Eq.~\eqref{eqw1} provides a different estimate of ${\cal
  Q}_\text{tot}(z_\text{tot})$ for each $i$, which ideally should be
combined to give a single best approximation.  The WHAM method
resolves this problem, yielding a best estimate for ${\cal
  Q}_\text{tot}(z_\text{tot})$ of the form:
\begin{equation}\label{eqw2}
{\cal Q}_\text{tot}(z_\text{tot}) = A \frac{\sum_{i=1}^M n_i {\cal P}_\text{tot}^{(i)}(z_\text{tot})}{\sum_{j=1}^M n_j e^{-\beta (U_j(z_\text{tot})-F_j)}},
\end{equation}
where $A$ is a normalization constant.  The unknown parameters
$\{F_i\}$ are given by:
\begin{equation}\label{eqw3}
F_i = -\frac{1}{\beta} \ln \left[ \int dz_\text{tot} {\cal Q}_\text{tot}(z_\text{tot}) e^{-\beta U_i(z_\text{tot})}\right].
\end{equation}
Eqs.~\eqref{eqw2} and \eqref{eqw3} are a coupled system of equations
for ${\cal Q}_\text{tot}(z_\text{tot})$ and $\{F_i\}$.  We solve these
by making an initial guess for the set $\{ F_i \}$, substituting it
into Eq.~\eqref{eqw2} to find ${\cal Q}_\text{tot}(z_\text{tot})$, and
using this estimate for ${\cal Q}_\text{tot}(z_\text{tot})$ in
Eq.~\eqref{eqw3} to find a new set of $\{ F_i \}$.  The process is
iterated until we converge to a self-consistent solution to both
equations.  Once we have a best estimate of ${\cal
  Q}_\text{tot}(z_\text{tot})$, we can calculate $\tilde{\cal
  P}_\text{tot}(z_\text{tot};F_0)$ as above, completing Step 1 of the
construction.

\section{Leucine zipper simulations}

\subsection{SOP model for the LZ26 leucine zipper}

The amino acid sequence for a single $\alpha$-helical strand of the LZ26
coiled coil is as follows (grouped into heptad repeats): MCQLEQK
VEELLQK NYHLEQE VARLKQL VGELEQK VEELLQK NYHLEQE VARLKQL VGELEQK
VEELLQK NYHLEQE VARLKQL VGEC.  The sequence is the same as in
Ref.~\cite{Gebhardt10}, except that we have left out four residues at
the beginning and three from the final heptad, for a total of 88
residues per strand. As in the experiment~\cite{Gebhardt10}, the
handles are attached at the cysteine in position b of the first
heptad, and the cross-linking between strands is at the cysteine in
position d of the last heptad.  (For consistency when comparing
simulations with or without the handles/traps, end-to-end distance for
the protein is always measured between the two N-terminal cysteines.)
Although the crystal structure is not available for LZ26, it is
believed to be similar to three GCN4 leucine zipper domains (PDB ID
code 2ZTA)~\cite{OShea91} in series.  Thus, we constructed a model for
the native structure based on GCN4, connecting the leucine zipper
segments in such a way that the distances between neighboring
$C_\alpha$ positions and angles of superhelical coiling formed a
continuous pattern as one moves along LZ26.

Going from N- to C-terminus on one strand and returning C- to
N-terminus on the other, let us label the residues
$i=1,\ldots,N_\text{res}$, where $N_\text{res}=176$, where $1\le i \le
88$ corresponds to one strand, and $88 < i \le 176$ corresponds to the
other.  Every non-neighboring pair of residues, $(i,j)$ where
$|i-j|>1$, is assigned to one of three sets: ${\cal S}$
(secondary structure pairs), ${\cal T}$ (tertiary structure pairs),
and ${\cal R}$ (remainder).  The set ${\cal S}$ consists of all pairs
where $|i-j| =4$ and $i$, $j$ share the same strand, representing
residues interacting through $\alpha$-helical hydrogen bonding.  The set
${\cal T}$ consists of all pairs where $i$, $j$ are on different
strands, and the distance between the two residues in the native
structure, $r^0_{i,j}$, is below a cutoff: $r^0_{i,j} < R_c=0.8$ nm.
These pairs are involved in tertiary interactions between the two
$\alpha$-helical coils.  All other non-neighboring pairs that do not
satisfy the criteria for ${\cal S}$ or ${\cal T}$ fall into the
set ${\cal R}$, and only interact via repulsive Lennard-Jones
potentials.  

The variant SOP Hamiltonian for LZ26 has the form:
\begin{equation}\label{ap1}
\begin{split}
{\cal H}_\text{SOP} =&
\frac{k_\text{bond}}{2}\sum_{i=1}^{N_\text{res}-1}
(r_{i,i+1}-r_{i,i+1}^0)^2 +
\frac{k_\text{ang}}{2}\sum_{i=1}^{N_\text{res}-2} \left[1-\cos(\theta_{i,i+1,i+2}-\theta_0)\right]\Delta_{i,i+1,i+2}\\
& - \sum_{(i,j) \in {\cal S}} \epsilon_\text{hb}\left[1+k^\text{hb}_\text{bond} (r_{i,j} - r_{i,j}^0)^2 + k^\text{hb}_\text{ang}\left( (\theta_{i,j,j-1} - \theta_{i,j,j-1}^0)^2 + (\theta_{i+1,i,j}-\theta_{i+1,i,j}^0)^2\right.\right.\\
&\qquad\qquad\qquad\left.\left.+ (\phi_{i+1,i,j,j-1} - \phi_{i+1,i,j,j-1}^0)^2\right)\right]^{-1}\\
& + \sum_{(i,j) \in {\cal T}} \chi |\epsilon_\text{BT}(i,j)-\epsilon_s| V_\text{LJ}\left(\frac{r^0_{i,j}}{r_{i,j}}\right) + \sum_{(i,j) \in {\cal R}} \epsilon_\text{rep} \left(\frac{\sigma}{r_{i,j}}\right)^6.
\end{split}
\end{equation}
The first term is the nearest-neighbor bond potential, where $r_{i,j}$
is the distance between residues $i$ and $j$, and the spring constant
$k_\text{bond} = 200$ kcal/mol$\cdot$nm$^2$.  The second term is the
bond angle potential, with the spring constant $k_\text{ang} = 2$
kcal/mol.  The angle between the bonds $(j,i)$ and $(j,k)$ is
$\theta_{i,j,k}$, and the equilibrium value $\theta_0 = 0.583\pi\:\text{rad} = 105^\circ$, a
typical bond angle in protein structures~\cite{Klimov98}.  The factor
$\Delta_{i,j,k} = 1$ if $i$, $j$, $k$ are all on the same strand, 0
otherwise.  The relative softness of the bond angle potential,
together with the form of the secondary structure interactions
detailed below, ensure that the two strands in the unfolded LZ26 (with
all inter-strand tertiary contacts broken) have a persistence length
of $\sim 0.7$ nm, consistent with experimental
measurements~\cite{Gebhardt10}.

The third term in Eq.~\eqref{ap1} accounts for the effects of hydrogen
bonding along the $\alpha$-helical backbone, and is based on a similar form developed for 
RNA~\cite{Denesyuk11}.  We mimic the directionality-dependence of
hydrogen bonds  by making the bond energy depend not only on the
distance $r_{i,j}$, but also on bond and dihedral angles defined by
the four residues $i+1$, $i$, $j$, and $j-1$, with $|i-j| = 4$.  For
each $(i,j)$ there are two bond angles, $\theta_{i,j,j-1}$ and
$\theta_{i+1,i,j}$, and one dihedral angle, $\phi_{i+1,i,j,j-1}$. The
equilibrium values of the angles, denoted by a superscript 0, are
calculated from the corresponding quantities in the native structure.
Only when the distance, bond angles, and dihedral angles are all
simultaneously equal to the equilibrium values does the hydrogen bond
potential reach its energy minimum $-\epsilon_\text{hb}$, where
$\epsilon_\text{hb}>0$.  Thus, the minimum is reached only when the
entire $(i,i+4)$ strand segment adopts a structure resembling a single
$\alpha$-helical turn.  The $\alpha$-helical propensity of an $(i,i+4)$
segment is determined by the energy scale $\epsilon_\text{hb}$ and the
sensitivity parameters $k_\text{bond}^\text{hb}$,
$k_\text{ang}^\text{hb}$.  Larger values for the sensitivity
parameters increase the brittleness of the $\alpha$-helix, making it more
likely to be destablized due to thermal fluctuations.  To calibrate the
parameters, we define a helix function $H(i,j)$ for any $(i,j) \in
{\cal S}$,
\begin{equation}\label{ap1b}
H(i,j) = \sqrt{\left(\frac{r_{i,j}}{r_{i,j}^0}-1\right)^2 + \left(\frac{\theta_{i,j,j-1}}{\theta_{i,j,j-1}^0}-1\right)^2  + \left(\frac{\theta_{i+1,i,j}}{\theta_{i+1,i,j}^0}-1\right)^2+ \left(\frac{\phi_{i+1,i,j,j-1}}{\phi_{i+1,i,j,j-1}^0} -1\right)^2 },
\end{equation}
reflecting the RMS deviation of the bond distances and angles from
their equilibrium values.  We use $H(i,j)$ as a measure of helix
content, by counting the fraction of pairs in ${\cal S}$ where
$H(i,j)$ is less than a cutoff $H_\text{c} = 0.5$.  It is known from
thermal denaturation experiments on GCN4~\cite{Holtzer01} that the
individual $\alpha$ helices upon unzipping are  unstable, with
$\approx 17\%$ helical content.  In contrast, the tertiary contacts in
the coiled-coil structure stabilize helix formation, resulting in a
much higher helical content of $\approx 81\%$.  We expect
qualitatively similar behavior in LZ26 in the case of force
denaturation, and thus tune the sensitivity parameters to yield a
large difference in the helix content between the unfolded and folded
states.  The parameter values are set at $\epsilon_\text{hb} = 3.85$
kcal/mol, $k^\text{hb}_\text{bond} = 10$ nm$^{-2}$,
$k^\text{hb}_\text{ang} = 40$ rad$^{-2}$.  For these values we find a
helix content of 3\% and 82\% respectively for the unfolded and folded
states of LZ26 under a constant force of $F_0 = 12.3$ pN.

The fourth term in Eq.~\eqref{ap1} describes tertiary interactions
between the two strands of LZ26, $(i,j) \in {\cal T}$.  These have a
residue-dependent energy $\chi |\epsilon_\text{BT}(i,j)-\epsilon_s|$.
Here $\chi$ is an overall prefactor, $\epsilon_\text{BT}(i,j)$ is the
Betancourt-Thirumalai (BT) contact energy for residues $i$ and
$j$~\cite{Betancourt99}, and $\epsilon_s$ shifts the zero of the
energy scale~\cite{Liu11}.  To get a leucine zipper that unfolds at
the experimental force scale of $\sim 12$ pN, we choose 
$\chi=2.25$ and energy shift $\epsilon_s = 0.7$ $k_B T$.  The tertiary
interactions use a modified Lennard-Jones potential of the form:
\begin{equation}\label{ap2}
V_\text{LJ}(x)=\begin{cases} x^6 - \frac{3}{2}x^4 -\frac{1}{2} & x \le 1\\ x^{12} - 2x^6 & x>1\end{cases}.
\end{equation}
This has the standard 12-6 form at large distances, but a softer
short-range repulsive core, increasing with the inverse 6th rather
than 12th power.  The choice of the softer potential is made to allow
for a longer simulation time step, while not having a significant
impact on the large-scale dynamics of the system~\cite{Hyeon06}.

The final term in Eq.~\eqref{ap1} describes purely repulsive
interactions among the remaining non-neighboring pairs, $(i,j) \in
{\cal R}$, with energy factor $\epsilon_\text{rep} = 1$ kcal/mol and
range $\sigma = 0.38$ nm.  We use the inverse 6th power in the
repulsive potential for the same reasons as above.

\subsection{Semiflexible bead-spring model for the DNA handles}

Each double-stranded DNA handle is modeled as a chain of
$N_\text{h}$ beads of radius $a = 1$ nm, corresponding to a contour
length $L=2aN_\text{h}$.  The handle Hamiltonian is:
\begin{equation}\label{ap3}
\begin{split}
{\cal H}_\text{h} =
\frac{k_\text{bond}}{2}\sum_{i=1}^{N_\text{h}-1}
(r_{i,i+1}-2a)^2 + \frac{l_p k_B T}{2a}\sum_{i=1}^{N_\text{h}-2} \left[1-\cos(\theta_{i,i+1,i+2})\right]
\end{split}
\end{equation}
where $r_{i,i+1}$ are the distances between neighboring beads,
$k_\text{bond} = 200$ kcal/mol$\cdot$nm$^2$, $l_p$ is the persistence
length, and $\theta_{i,i+1,i+2}$ are angles between consecutive bonds.
The two terms are stretching and bending energies respectively.  The
handle elastic modulus $\gamma = 2a k_\text{bond} = 2780$ pN. For the
persistence length we consider, $l_p = 20$ nm, at the applied tension
due to the traps, the handles (and unfolded portions of the protein)
are almost fully extended, and there is negligible probability of the
chain overlapping itself or protein residues in the vicinity of the
handle attachment point.  Hence, there is no need to include excluded
volume interactions for the handles.  The covalent linkers that attach
the handles either to the cysteine residue at the protein N-terminus
or a point on the bead surface are modeled as simple harmonic springs
with strength $\kappa = k_\text{bond}$ and length $\ell = 1.5$ nm.

\subsection{Simulation time scales}

Let $\mu_0 = 1/6\pi \eta a$ be the mobility of a sphere of radius $a =
1$ nm, where $\eta = 0.89$ mPa$\cdot$s is the viscosity of water at
$T=298$ K.  This will be the mobility of our DNA handle beads, while
for the large polystyrene beads the corresponding mobility is
$\mu_\text{b} = \mu_0a/R_b$.  The rotational diffusion of the
polystyrene bead is characterized by a mobility $\mu_\text{b}^r =
3\mu_0a/4R_b^3$.  For the protein residues we choose a mobility
$\mu_\text{res} = 3.36\mu_0$, corresponding to an effective
hydrodynamic radius of $0.30$ nm.  The characteristic Brownian
dynamics time scale associated with $\mu_0$ is $\tau_0 = a^2/\mu_0 k_B
T = 4$ ns.  To avoid numerical errors, our simulation time step $\tau$
should be a small fraction of $\tau_0$, and we obtained reliable
results using $\tau = 5\times 10^{-6}\tau_0 = 0.02$ ps.  For LZ26 both
with and without handles/beads, we ran $\approx 260$ long trajectories
at various force conditions (or trap separations), totaling to
$\approx 10^{12}$ simulation time steps, or $20$ ms, with
data collection every $10^4$ steps.  (In the case of the
simulations involving the GRM hairpin instead of the protein, the time
step $\tau = 3\times 10^{-4}\tau_0 = 1.2$ ps, and the total trajectory
data for each GRM parameter set corresponded to $160 - 180$ ms.)

\section{Finite bandwidth scaling (FBS):  correcting for the effects of electronic filtering, time averaging, and noise}\label{sec:fbs}

Before the data from optical tweezer experiments can be used to
reconstruct the intrinsic biomolecule free energy landscape, one must
consider the inevitable distortions due to noise, the electronic
systems involved in data recording, and any additional filtering done
as part of the collection protocol.  We have developed a method,
finite bandwidth scaling (FBS), to correct for these
distortions. In the following we first derive the basic FBS scaling
relations, and then verify them using both simulation and experimental
data sets.

\subsection{FBS theory}

Understanding how the time series of bead positions is distorted as
part of the measurement process requires a detailed spectral analysis
of all components in the dual optical tweezer apparatus.  The spectral
properties of the experimental system used to collect the data in our
work have been extensively characterized by von Hansen {\em
  et. al.}~\cite{vonHansen12}, allowing us to develop a simplified
theory which approximates the most important sources of distortion.
Our theory fits all the experimental data sets under consideration,
but it can be easily modified to include additional complications that
we ignore (for example crosstalk between the two laser traps) as well
as the details of other experimental setups.

Let $z^\text{raw}_\text{tot}(t)$ be the trajectory of bead-bead separations along
the $\hmb{z}$-axis recorded during the experiment.  This raw data set
is based on the signal from the silicon photodiode devices that
measure the deflection of the lasers due to bead displacements.  This
output is then processed and amplified by the electronic system used
in the recording apparatus.  If $z_\text{tot}(t)$ is the actual trajectory of
bead displacements, inaccessible to the experimentalist, the recorded
output $z^\text{raw}_\text{tot}(t)$ is related to $z_\text{tot}(t)$ as:
\begin{equation}\label{f1}
z^\text{raw}_\text{tot}(t) = \int_{-\infty}^t f(t-t^\prime)(z_\text{tot}(t^\prime)+\eta(t^\prime)).
\end{equation}
The deviation of $z^\text{raw}_\text{tot}(t)$ from $z_\text{tot}(t)$
stems from two main effects: (i) an additive noise component
$\eta(t)$, which includes environmental noise like vibrations of the
optical elements in the apparatus and electronic noise in the
detectors~\cite{vonHansen12}.  For simplicity, we model the noise as
Gaussian white noise with zero mean and variance equal to $\nu$:
$\langle \eta(t) \rangle =0$, $\langle \eta(t)\eta(t^\prime) \rangle =
\nu \delta(t-t^\prime)$, and $\langle \eta(t) z_\text{tot}(t^\prime)
\rangle = 0$, where $\langle \; \rangle$ denotes an equilibrium
ensemble average; (ii) convolution with a kernel function $f(t)$,
which reflects the filtering properties of the photodiodes and
electronics.  Any additional time averaging or filtering carried out
by the experimentalist on the recorded data series will be considered
explicitly later on, and is not included in $f(t)$.  The analysis of
Ref.~\cite{vonHansen12} yielded the following form for the filter
kernel in the frequency domain,
\begin{equation}\label{f2}
f(\omega) = \left[\lambda +  \frac{1-\lambda}{1 -i \omega \tau_{pf}} \right]\frac{1}{B_8(i \omega \tau_{bf})},
\end{equation}
where for our LOT setup $\lambda=0.6\pm 0.05$, $\tau_{pf} = 6 \pm 1$
$\mu$s, $\tau_{bf} = 5$ $\mu$s, and $B_8(x)$ is the 8th order
Butterworth polynomial.  The term in the square brackets above
originates in a physical phenomenon known as ``parasitic
filtering''~\cite{Peterman03,Berg06}, arising from the transparency of
the silicon in the photodiode to the laser light with wavelength 1064
nm used in the experiment: a fraction $1-\lambda$ of the photocurrent
from the detector is produced with a lag time $\tau_{pf}$ relative to
the photon signal.  The second term in Eq.~\eqref{f2}, involving the
Butterworth polynomial, is due to the subsequent electronic
amplification of the signal from the detector, which acts like a
Butterworth lowpass filter with characteristic timescale $\tau_{bf}$,
such that at the frequency $\omega = \tau_{bf}^{-1}$ the signal
amplitude is attenuated by 3 dB.  Since the form of Eq.~\eqref{f2} is
too complicated for use in our analytical theory, we will approximate
$f(t)$ as a generic first-order low-pass filter, exploiting the fact
that both the parasitic and electronic terms act to attenuate
high-frequency portions of the signal,
\begin{equation}\label{f3}
f(\omega) \approx \frac{1}{1-i \omega \tau_{f}},
\end{equation}
where $\tau_{f}=7$ $\mu$s.  This effective filtering timescale
$\tau_f$ is derived by demanding that Eq.~\eqref{f3} exhibit the same
degree of attenuation at $\omega = \tau_{bf}^{-1}$ as Eq.~\eqref{f2}.

Though these distortions are expressed in the frequency domain, they
have observable consequences for the equilibrium probability
distribution of bead-bead separations.  As an example, consider
the raw autocorrelation function $C_\text{raw}(t) = \langle
(z^\text{raw}_\text{tot}(t)-\bar{z}^\text{raw}_\text{tot})(z^\text{raw}_\text{tot}(0)-\bar{z}^\text{raw}_\text{tot})\rangle$,
where $\bar{z}^\text{raw}_\text{tot}$ is the mean recorded bead-bead
separation.  The variance of the raw probability distribution ${\cal
  P}^\text{raw}_\text{tot}(z^\text{raw}_\text{tot})$ is equal to
$C_\text{raw}(0)$.  From Eqs.~\eqref{f1} and \eqref{f3}, the raw
autocorrelation is related to the true one, $C(t) = \langle
(z_\text{tot}(t)-\bar{z}_\text{tot})(z_\text{tot}(0)-\bar{z}_\text{tot})\rangle$,
by:
\begin{equation}\label{f4}
C_\text{raw}(t) = \frac{\nu}{2 \tau_{f}} e^{-t/\tau_{f}} + \int_{-\infty}^\infty dt^\prime\, \frac{e^{-|t-t^\prime|/\tau_f}}{2\tau_f} C(t^\prime).
\end{equation}

The first term in Eq.~\eqref{f4}, due to noise, tends to increase the
variance $C_\text{raw}(0)$ relative to $C(0)$.  The second term, due
to filtering, is always less than $C(0)$, since it is an average over
$C(t^\prime)$, and $C(t^\prime \ne 0) < C(0)$.  Noise broadens
the measured distribution, and filtering narrows it.  However without
knowing the amplitude of the noise $\nu$, it is unclear whether the
filtering due to the detectors and electronics under- or
over-compensates for the noise, and how far ${\cal
  P}^\text{raw}_\text{tot}(z^\text{raw}_\text{tot})$ deviates from the
true distribution ${\cal P}_\text{tot}(z_\text{tot})$.  Thus, we need a
way to estimate $\nu$.

The situation is even more complicated since the experimentalist may
choose to apply additional filtering on the recorded data, for example
as a way of manually removing noise and unwanted high frequency
components of the signal (since the dynamics of interest typically
occur at frequencies much lower than imposed filter cutoff).  For the
GCN4 leucine zipper, the data sets recorded at 100 kHz (corresponding
to a sampling time step $\tau_s = 10$ $\mu$s) were subsequently
filtered in real time during collection by averaging every 5
consecutive time steps together.  Such averaging acts like a low-pass
filter, and so has narrowing effects on the equilibrium probability
distribution qualitatively similar to the filtering described above.
Some type of additional filtering of this kind is a common
experimental practice (see Refs.~\cite{Gao11},\cite{Yu12}, and
\cite{Ritchie12} for recent examples, involving either an averaging or 8
pole Bessel filter).  It turns out, however, that we can take
advantage of the filtering protocol: by varying the degree of
filtering we will use it to approximately extrapolate features of the
true probability distribution.

Let us concentrate on the simple case of filtering the recorded data
by averaging every $n$ consecutive points into a single value.  If the
collection time step is $\tau_s$, the original raw data is represented
by the recorded time series $\{ z^\text{raw}_\text{tot}(t_j)\}$, where
$t_j \equiv j \tau_s$ for $j=1,2,3,\ldots$.  The averaged data is a
time series $\{ z^\text{raw,$n$}_\text{tot}(t_{nj})\}$, where
$z^\text{raw,$n$}_\text{tot}(t_{nj}) = n^{-1} \sum_{i=nj-n+1}^{i=nj}
z^\text{raw}_\text{tot}(t_i)$.  For the averaged time series we will
focus on two quantities, both related directly to its autocorrelation
$C_\text{raw,$n$}(t)$: the variance $\langle
(z^\text{raw,$n$}_\text{tot}-\bar{z}^\text{raw,$n$}_\text{tot})^2\rangle
= C_\text{raw,$n$}(0)$, and the mean-squared displacement (MSD)
between consecutive points, $\langle (z^\text{raw,$n$}_\text{tot}(n
\tau_s)- z^\text{raw,$n$}_\text{tot}(0))^2 \rangle =
2(C_\text{raw,$n$}(0)-C_\text{raw,$n$}(n\tau_s)) \equiv
\Delta_\text{raw,$n$}(n\tau_s)$.  In a more complicated fashion, these
two quantities can also be expressed in terms of the original autocorrelation
$C_\text{raw}(t) = C_\text{raw,1}(t)$ before averaging:
\begin{equation}\label{f5}
\begin{split}
C_\text{raw,$n$}(0) &= \frac{1}{n}C_\text{raw}(0) +\frac{2}{ n^2}\sum_{j=1}^{n-1}(n-j)C_\text{raw}(j\tau_s),\\
\Delta_\text{raw,$n$}(n\tau_s) &=\frac{2}{n}C_\text{raw}(0)+ \frac{2}{n^2} \sum _{j=1}^n (2 n-3 j)C_\text{raw}(j\tau_s) -\frac{2}{n^2} \sum _{j=1}^{n-1} (n-j) C_\text{raw}((n+j)\tau_s).
\end{split}
\end{equation}
We know that $C_\text{raw}(t)$ is related to the unknown true
correlation $C(t)$ through Eq.~\eqref{f4}, so we can complete the
theoretical description by specifying a form for $C(t)$.  A generic
correlation function can be expanded as a sum of exponentials, $C(t) =
\sum_{i=1}^\infty A_i \exp(-t/\tau_i)$, with relaxation times
$\tau_1<\tau_2<\cdots$.  We will be interested in correlations on the
shortest accessible time-scales, $t \sim {\cal O}(\tau_s)$, so we plug
the expression for $C(t)$ into Eq.~\eqref{f4} and expand for small
$t$, keeping the contribution from the $\tau_1$ exponential and lowest
order corrections from the $\tau_{i>1}$ terms:
\begin{equation}\label{f6}
\begin{split}
C_\text{raw}(t) &= \frac{\nu}{2 \tau_{f}} e^{-t/\tau_{f}} + \sum_{i=1}^\infty \frac{A_i \tau_i}{\tau_i^2-\tau_f^2}\left(\tau_i e^{-t/\tau_i}-\tau_f e^{-t/\tau_f}\right)\\
&\approx \frac{\nu}{2 \tau_{f}} e^{-t/\tau_{f}} + \frac{A_1 \tau_i}{\tau_1^2-\tau_f^2}\left(\tau_1 e^{-t/\tau_1}-\tau_f e^{-t/\tau_f}\right) + A_c - B_c (t+ t_f e^{-t/\tau_f}),
\end{split}
\end{equation}
where $A_c = \sum_{i=2}^\infty A_i$, $B_c = A_2/\tau_2$.  If
necessary, the expansion can be extended to higher orders, but the
above form was sufficient to fit all the simulation and experimental
cases which we analyze below.

Eqs.~\eqref{f5}-\eqref{f6} completely define the variance
$C_\text{raw,$n$}(0)$ and MSD $\Delta_\text{raw,$n$}(n\tau_s)$ in
terms of five unknown parameters: $\nu$, $A_1$, $\tau_1$, $A_c$, and
$B_c$.  By averaging the recorded time series $\{
z^\text{raw}_\text{tot}(t_j)\}$ for different values of $n$ (varying
the effective filter bandwidth), we construct curves of
$C_\text{raw,$n$}(0)$ and MSD $\Delta_\text{raw,$n$}(n\tau_s)$ as a
function of $n$.  Fitting these curves to Eqs.~\eqref{f5}-\eqref{f6},
we can then extract the unknown parameters.  This allows us to
estimate the true variance of the probability distribution, 
\begin{equation}\label{f6b}
C(0) = A_1 + A_c.
\end{equation}
Since we are using properties of time series at different
effective bandwidths to gain information about the true, ``infinite''
bandwidth limit, we call our method {\it finite bandwidth scaling}
(FBS).  The analogy is to finite size scaling~\cite{Ferdinand69},
where thermodynamic properties of systems on finite lattices are
extrapolated to the infinite lattice limit.  One of the nice features
of FBS is that the scaling analysis can be carried out even when we
can only calculate $C_\text{raw,$n$}(0)$ and
$\Delta_\text{raw,$n$}(n\tau_s)$ for a subset of $n$ values.  For
example, in the leucine zipper case below, the available time series
corresponds to $n=5$, since the data was time averaged during
collection.  From the $n=5$ data we can construct trajectories for
$n=10,15,20,\ldots$.  This subset is sufficient for the FBS
extrapolation.

Once we know $C(0)$, how can we use it to approximately reconstruct
the true distribution ${\cal P}_\text{tot}(z_\text{tot})$?  Keep in
mind that the variance $C(0) = \langle z_\text{tot}^2 \rangle -
\langle z_\text{tot} \rangle^2 = \int dz\, (z - \langle
z_\text{tot}\rangle)^2 {\cal P}_\text{tot}(z)$.  The simplest estimate
for ${\cal P}_\text{tot}(z_\text{tot})$ is to start with the measured,
averaged distribution ${\cal P}^\text{raw,$n$}_\text{tot}$ for some
$n$, and deform it in one of two ways, changing its variance by an
amount $\delta_C = |C(0) - C_\text{raw,$n$}(0)|$: (i) If
$C_\text{raw,$n$}(0) < C(0)$, we carry out a convolution with a
normalized Gaussian of variance $\delta_C$,
\begin{equation}\label{f7}
{\cal P}_\text{tot}(z_\text{tot}) \approx \int_{-\infty}^\infty dz\, {\cal P}^\text{raw,$n$}_\text{tot}(z_\text{tot}-z) \frac{e^{-z^2/2\delta_c}}{\sqrt{2\pi \delta_C}}.
\end{equation}
(ii) If $C_\text{raw,$n$}(0) > C(0)$, we do a deconvolution instead, solving
\begin{equation}\label{f7}
{\cal P}^\text{raw,$n$}_\text{tot}(z_\text{tot}) \approx \int_{-\infty}^\infty dz\, {\cal P}_\text{tot}(z_\text{tot}-z) \frac{e^{-z^2/2\delta_c}}{\sqrt{2\pi \delta_C}}
\end{equation}
for ${\cal P}_\text{tot}$.  The latter can be carried out using the
numerical deconvolution technique described in Sec.~\ref{sec:numdec}.
After the deformation, the estimated ${\cal P}_\text{tot}$ will by
construction have the correct variance $C(0)$.  We should recover
roughly the same ${\cal P}_\text{tot}$ starting from ${\cal
  P}^\text{raw,$n$}_\text{tot}$ for any $n$ in the range where the FBS
scaling is valid, as we will demonstrate in the examples below.  In
systems with multiple states, where there is more than one peak in the
measured distribution, it is more accurate to carry out the FBS
analysis separately on each state, and apply the corresponding
specific deformation for each peak.  This can be done with the aid of
hidden Markov model~\cite{Rabiner89} partitioning of the time series,
as described in the next section for the case of the GRM and leucine
zipper.

The FBS method has several limitations: (i) using a Gaussian to deform
${\cal P}^\text{raw,$n$}_\text{tot}$ into ${\cal P}_\text{tot}$ is an
assumption, since all we strictly know about the actual point-spread
function is the variance $\delta_C$.  The smaller the variance, the
more valid the assumption, since the potential non-Gaussian
contributions to the point-spread function become less significant.
We can also test the assumption from our measured data, by checking
whether the ${\cal P}^\text{raw,$n$}_\text{tot}$ for various $n$ can
be mapped to each other by Gaussian deformations.  For all the systems
analyzed in our work this is indeed the case.  (ii) Gaussian
deformations map individual peaks into slightly broader or narrower
peaks, but do not produce new peaks.  Hence, if there is a state with
a very short lifetime that is smeared out by the filtering (either the
parasitic, electronic, or additional filtering), yielding no distinct
peak in ${\cal P}^\text{raw,$n$}_\text{tot}$, the FBS method will not
be able to reconstruct its properties.  Whether or not the
experimentalist chooses to do additional averaging, the intrinsic time
resolution $\tau_f$ of the apparatus puts fundamental constraints on
what we can learn from the measured time series.  Transitions
occurring on timescales faster than $\tau_f$ will be lost to us.
(iii) In a similar way, the characteristic relaxation times of the
trapped beads also impose limits.  To illustrate this, take two
protein states $S_1$ and $S_2$, which have a small difference in their
mean end-to-end distance along the force direction, and assume $S_1$
is only accessible from $S_2$.  If the mean lifetime of $S_1$ is much
smaller than $S_2$, such that it is shorter than the bead relaxation
time, transitions like $S_2 \to S_1 \to S_2$ will correspond to only
negligible excursions in the measured trajectory of bead
displacements, since the beads do not have enough time to relax to the
equilibrium position associated with $S_1$ before the protein returns
to $S_2$.  If the mean-squared distance of the excursions is smaller
than the noise amplitude in the recorded time series, the existence of
state $S_1$ will be hidden from the experimentalist, regardless of the
apparatus filtering timescale $\tau_f$.  In summary, the distribution
produced by FBS is an approximation to the truth: the method can
correct distortions produced by noise and filtering, but it only works
for states in the energy landscape which leave some signature of
themselves in the measured time series.

\setlength{\tabcolsep}{10pt}
\begin{table}[h]
  \caption[FBS parameters]{Parameters used in the FBS analysis of simulation and experimental systems}\label{tab1}
\vspace{0.5em}
\centering \begin{tabular}{|lccccccc|}
\hline
& $\tau_s$ & $\tau_f$ & $\nu$ & $A_1$ & $\tau_1$ & $A_c$ & $B_c$\\
& [$\mu$s] & [$\mu$s] &  [nm$^2\mu$s] & [nm$^2$] & [$\mu$s] & [nm$^2$] & [nm$^2$/$\mu$s]\\
\hline
\multicolumn{8}{|l|}{Simulation:  GRM}\\
\hline
State $N$ & 0.024 & 0 & 0 & $2.2 \pm 1.2$ & $3.2 \pm 1.0$ & $3.1\pm 1.2$ & $0.24 \pm 0.15$\\
State $U$ & 0.024 & 0 & 0 & $1.6 \pm 0.2$ & $2.1 \pm 0.2$ & $2.6 \pm 0.2$ & $0.22 \pm 0.04$\\
\hline
\multicolumn{8}{|l|}{Experiment:  dsDNA handles (no protein)}\\
\hline
$F_0 = 9.4$ pN & 10 & 7 & $31.8 \pm 1.5$ & $5.0 \pm 0.2$ & $14.7 \pm 1.8$ & $3.2 \pm 0.2$ & $0.0077 \pm 0.0020$\\
$F_0 = 11.5$ pN & 10 & 7 & $31.8 \pm 1.5$ & $4.0 \pm 0.2$ & $12.6 \pm 1.4$ & $3.0 \pm 0.1$ & $0.0079 \pm 0.0012$\\
$F_0 = 12.7$ pN & 10 & 7 & $31.8 \pm 1.5$ & $3.3 \pm 0.2$ & $11.5 \pm 1.2$ & $2.9 \pm 0.1$ & $0.0076 \pm 0.0008$\\
\hline
\multicolumn{8}{|l|}{Experiment:  GCN4 leucine zipper (trajectory 1)}\\
\hline
State I1 & 10 & 7 & $31.8 \pm 1.5$ & $3.9 \pm 0.2$ & $28.1 \pm 5.0$ & $3.5 \pm 0.2$ & $0.0066 \pm 0.0010$\\
State I2 & 10 & 7 & $31.8 \pm 1.5$ & $8.6 \pm 0.6$ & $41.9 \pm 8.0$ & $4.2 \pm 0.8$ & $0.0069 \pm 0.0036$\\
State U   & 10 & 7 & $31.8 \pm 1.5$ & $7.6 \pm 0.1$ & $23.7 \pm 1.7$ & $3.2 \pm 0.1$ & $0.0065 \pm 0.0007$\\
\hline
\multicolumn{8}{|l|}{Experiment:  GCN4 leucine zipper (trajectory 2)}\\
\hline
State I1 & 10 & 7 & $31.8 \pm 1.5$ & $4.1 \pm 0.1$ & $28.1 \pm 3.5$ & $3.5 \pm 0.2$ & $0.0069 \pm 0.0008$\\
State I2 & 10 & 7 & $31.8 \pm 1.5$ & $8.3 \pm 1.0$ & $41.3 \pm 12.0$ & $5.3 \pm 1.3$ & $0.0074 \pm 0.0053$\\
State U   & 10 & 7 & $31.8 \pm 1.5$ & $8.1 \pm 0.2$ & $23.6 \pm 2.2$ & $3.2 \pm 0.2$ & $0.0066 \pm 0.0009$\\
\hline
\end{tabular}
\end{table}

\subsection{Testing FBS on simulation and experimental data}

\begin{figure}[t]
\centering\includegraphics*[scale=1]{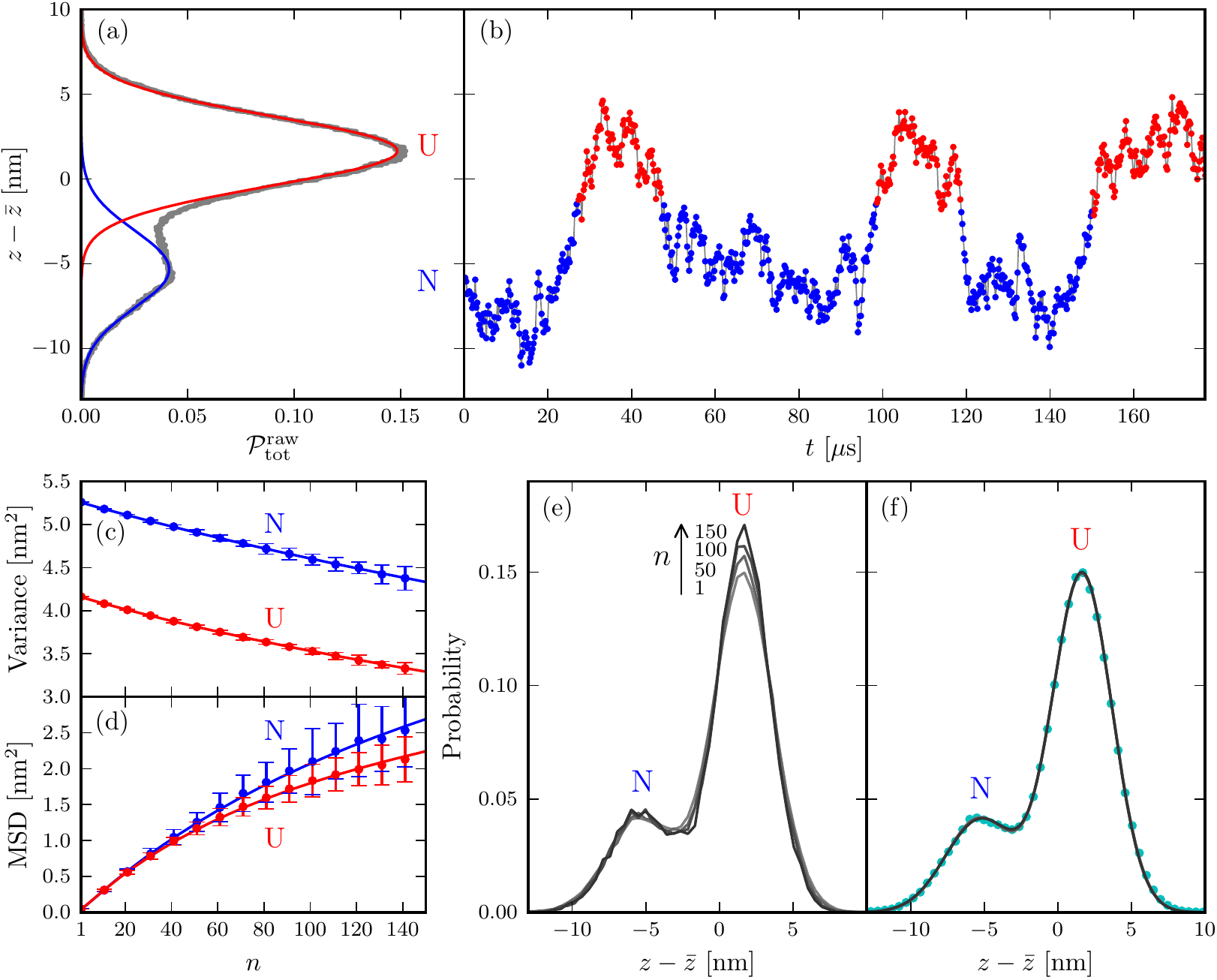}
  \caption{FBS analysis of a GRM Brownian dynamics simulation
    ($z_\text{trap} = 1298$ nm; all other parameters as in Table~S1).
    (a) The probability distribution of the bead-bead separation from
    the raw simulation data, ${\cal P}^\text{raw}_\text{tot}$ (gray),
    and the decomposition into individual Gaussian peaks corresponding
    to the N (blue) and U (red) states.  Distances are measured with
    respect to $\bar{z}$, the mean bead-bead separation.  (b) A sample
    time series fragment from the simulation, with the individual data
    points colored according to their assignment to the N (blue) and U
    (red) states by hidden Markov model analysis.  (c) For the raw
    time series filtered by averaging together every $n$ data points,
    the variance $C_\text{raw,$n$}(0)$ as a function of $n$.  The time
    series corresponding to each state, N and U, is analyzed
    separately, and plotted as blue and red points respectively, with
    bars denoting standard error due to finite sampling.  The solid
    curves are the FBS theoretical fits to $C_\text{raw,$n$}(0)$
    [Eq.~\eqref{f5}]. Best-fit FBS parameters are listed in
    Table~S2. (d) Analogous to (c), except showing the MSD function
    $\Delta_\text{raw,$n$}(n\tau_s)$ for consecutive pairs of points
    in the averaged time series.  (e) The raw distributions ${\cal
      P}^\text{raw,$n$}_\text{tot}$ for the averaged time series at
    $n=1$, 50, 100, 150.  (f) Solid curves: the distribution ${\cal
      P}_\text{tot}$ estimated by applying the appropriate FBS
    correction to the raw distributions in (e).  There are four
    curves, but due to overlapping they appear as one.  Points: the
    raw distribution ${\cal P}^\text{raw,$n$}_\text{tot}$ for $n=1$
    (no averaging), which for the GRM case is the true distribution,
    since there are no noise or apparatus filtering effects in the
    simulation.}\label{grm_avg}
\end{figure}

As a first test of the FBS theory, we analyze a Brownian dynamics
simulation trajectory of the GRM model in an optical tweezer setup
(Fig.~\ref{grm_avg}).  The trap separation $z_\text{trap} = 1298$ nm,
and all the other parameters are listed in Table~S1.
A computer simulation has perfect recording of data, with no
environmental noise or apparatus filtering effects, hence it can test
the FBS theory of Eqs.~\eqref{f5}-\eqref{f6} in the limit $\nu =
\tau_f =0$.  In this case the true distribution is just the $n=1$ raw
distribution ${\cal P}^\text{raw}_\text{tot} = {\cal
  P}^\text{raw,1}_\text{tot}$, plotted in Fig.~\ref{grm_avg}(a) (gray
curve).  If the FBS scaling is valid, we should be able to map any
distribution for $n>1$ onto the $n=1$ result by applying the FBS
correction procedure described above.

The GRM model exhibits two states, native N and unfolded U, which have
distinct dynamical properties.  Hence it is more accurate to apply the
FBS method separately to just those portions of the time series
belonging to each state.  Partitioning the time series by state
requires estimating the most likely sequence of states that
corresponds to the data.  Hidden Markov modeling
(HMM)~\cite{Rabiner89} is a general tool for this task.  The
probability distribution can be accurately decomposed into Gaussians
corresponding to each state, as depicted in Fig.~\ref{grm_avg}(a),
which define likelihoods for any observed $z_\text{tot}$ data point in
the trajectory to belong to one or the other state.  We then process
the entire trajectory through the Baum-Welch algorithm~\cite{Baum70},
to find optimal values for the unknown transition probabilities
between states, and finally construct the most likely state sequence
using the Viterbi algorithm~\cite{Viterbi67}.  Fig.~\ref{grm_avg}(b)
shows a fragment of the trajectory, colored according to the state
assignment resulting from HMM.

The variance $C_\text{raw,$n$}(0)$ and MSD
$\Delta_\text{raw,$n$}(n\tau_s)$ are then calculated as a function of
$n$ from the trajectory fragments belonging to a certain state.  For a
given $n$, the calculation involves averaging over data points within
a time window up to $2n\tau_s$ in length, so getting good statistics
requires having many fragments longer than $2n\tau_s$.  This will be
true so long as $2n\tau_s$ is much smaller than the mean lifetime of
the state, putting a practical upper bound on $n$.
Fig.~\ref{grm_avg}(c) and (d) plot the results for
$C_\text{raw,$n$}(0)$ and $\Delta_\text{raw,$n$}(n\tau_s)$
respectively (blue points: N state; red points: U state).  Bars
represent statistical standard error due to finite sample size, as
determined through jackknife estimation~\cite{Miller74}.  The solid
curves are fits to Eqs.~\eqref{f5}-\eqref{f6}, from which we extract
the FBS parameter values listed in Table~S2.  

With these values in hand, we can carry out the correction procedure:
Fig.~\ref{grm_avg}(d) shows the raw distributions ${\cal
  P}^\text{raw,$n$}_\text{tot}$ for $n=1,50,100,150$ (solid curves),
and panel (e) shows that same distributions after they have been
corrected according to the method outlined above ($n=1$ needs no
correction, but is included for comparison).  The greater the degree
of averaging (increasing $n$), the narrower the peaks in ${\cal
  P}^\text{raw,$n$}_\text{tot}$.  However, the FBS method compensates
for this, and all the distributions in (e) have collapsed onto a
single estimate for the true ${\cal P}_\text{tot}$.  As expected, this
estimate agrees very well with the $n=1$ result ${\cal
  P}^\text{raw}_\text{tot}$ (cyan points).

\begin{figure}[t]
\centering\includegraphics*[scale=1]{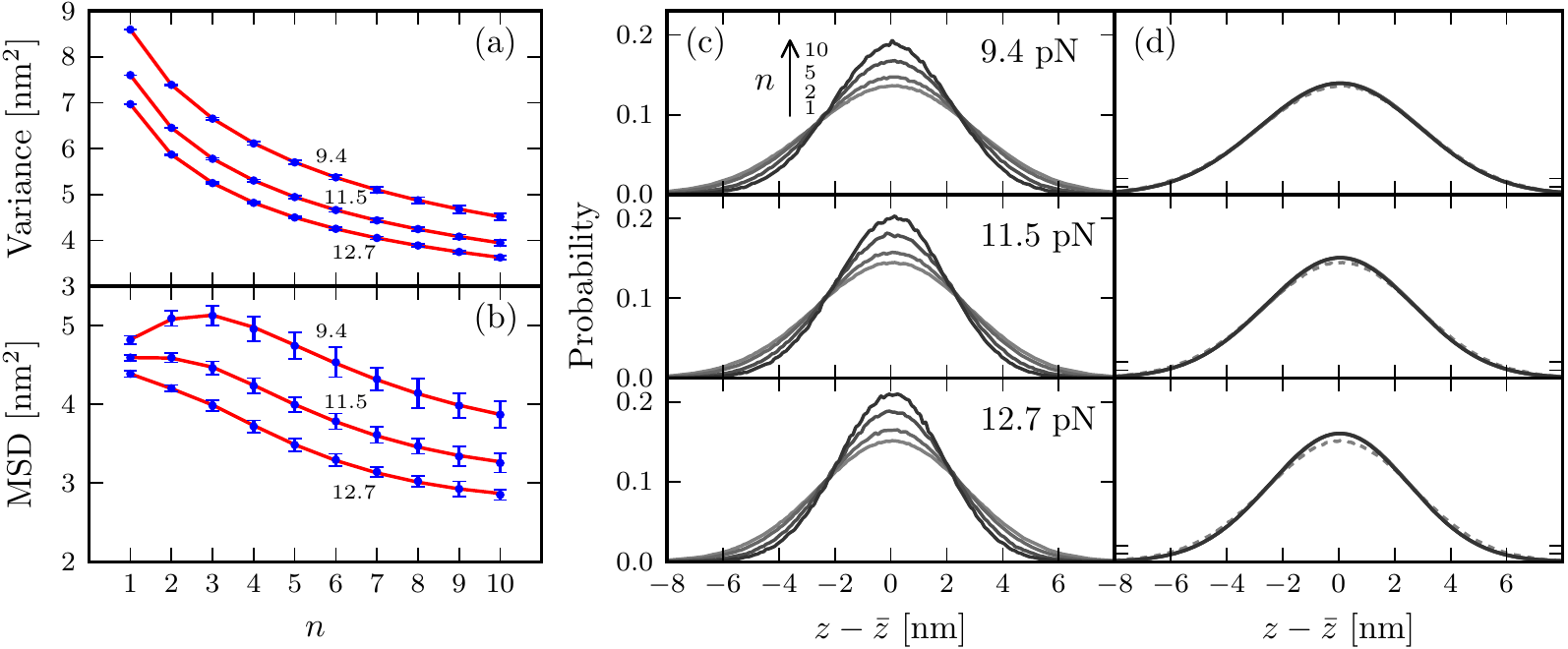}
\caption{FBS analysis of experimental results for a system containing
  only dsDNA handles and beads (see Sec.~\ref{sec:ps} for apparatus
  parameters).  We analyze three separate trajectories at different
  constant trap separations, corresponding to mean forces $F_0 =
  9.4\pm 0.7$, $11.5\pm 0.8$, and $12.7\pm 0.9$ pN.  The results in
  each panel are labeled by the $F_0$ value of the trajectory.  (a)
  For the raw experimental time series filtered by averaging together
  every $n$ data points, the variance $C_\text{raw,$n$}(0)$ as a
  function of $n$.  The results are plotted as points, with bars
  denoting standard error due to finite sampling.  The solid curves
  are the FBS theoretical fits to $C_\text{raw,$n$}(0)$
  [Eq.~\eqref{f5}]. Best-fit FBS parameters are listed in Table~S2.
  (b) Analogous to (a), except showing the MSD function
  $\Delta_\text{raw,$n$}(n\tau_s)$ for consecutive pairs of points in
  the averaged time series.  (c) The raw distributions ${\cal
    P}^\text{raw,$n$}_\text{tot}$ for the averaged time series at
  $n=1$, 2, 5, 10.  Each row corresponds to a different trajectory.
  (d) Solid curves: the distribution ${\cal P}_\text{tot}$ estimated
  by applying the appropriate FBS correction to the raw distributions
  in (c).  For each trajectory there are four curves, but due to
  overlapping they appear as one.  Dashed curves: the raw distribution
  ${\cal P}^\text{raw,$n$}_\text{tot}$ for $n=1$.  Though this
  distribution is free of any additional time averaging carried out on
  the recorded time series, it is subject to parasitic and electronic
  filtering effects intrinsic to the apparatus.  These distortions are
  corrected by FBS, and hence the dashed and solid curves are
  distinct.}\label{hand_avg}
\end{figure}

The second test of the FBS theory is on experimental data for a system
with only dsDNA handles and beads, discussed in Sec.~\ref{sec:ps}.
FBS results for three different trajectories (corresponding to three
values of the mean force $F_0$) are presented in Fig.~\ref{hand_avg}.
These data sets were recorded with a sampling rate of 100 kHz ($\tau_s
= 10$ $\mu$s), with no additional averaging beyond the unavoidable
filtering effects of the detectors and electronics.  As a consequence
of these effects, ${\cal P}^\text{raw,1}_\text{tot}$ is not the same
as the true distribution, and the deviation grows larger as $n$ is
increased.  The FBS best-fit results are shown in Table~S2.  In the
fitting the noise amplitude $\nu$ is constrained to be the same among
all three trajectories, since they are all collected on the same
equipment.  Like in the previous example, ${\cal
  P}^\text{raw,$n$}_\text{tot}$ for various $n$ can all be collapsed
onto a single estimate for ${\cal P}_\text{tot}$ through the FBS
method.  In Fig.~\ref{hand_avg}(e) this estimate (solid curves) is
compared to ${\cal P}^\text{raw,1}_\text{tot}$ (dashed curves), to
emphasize that the distortions due to apparatus filtering are small
but noticeable.

\begin{figure}[t]
\centering\includegraphics*[scale=1]{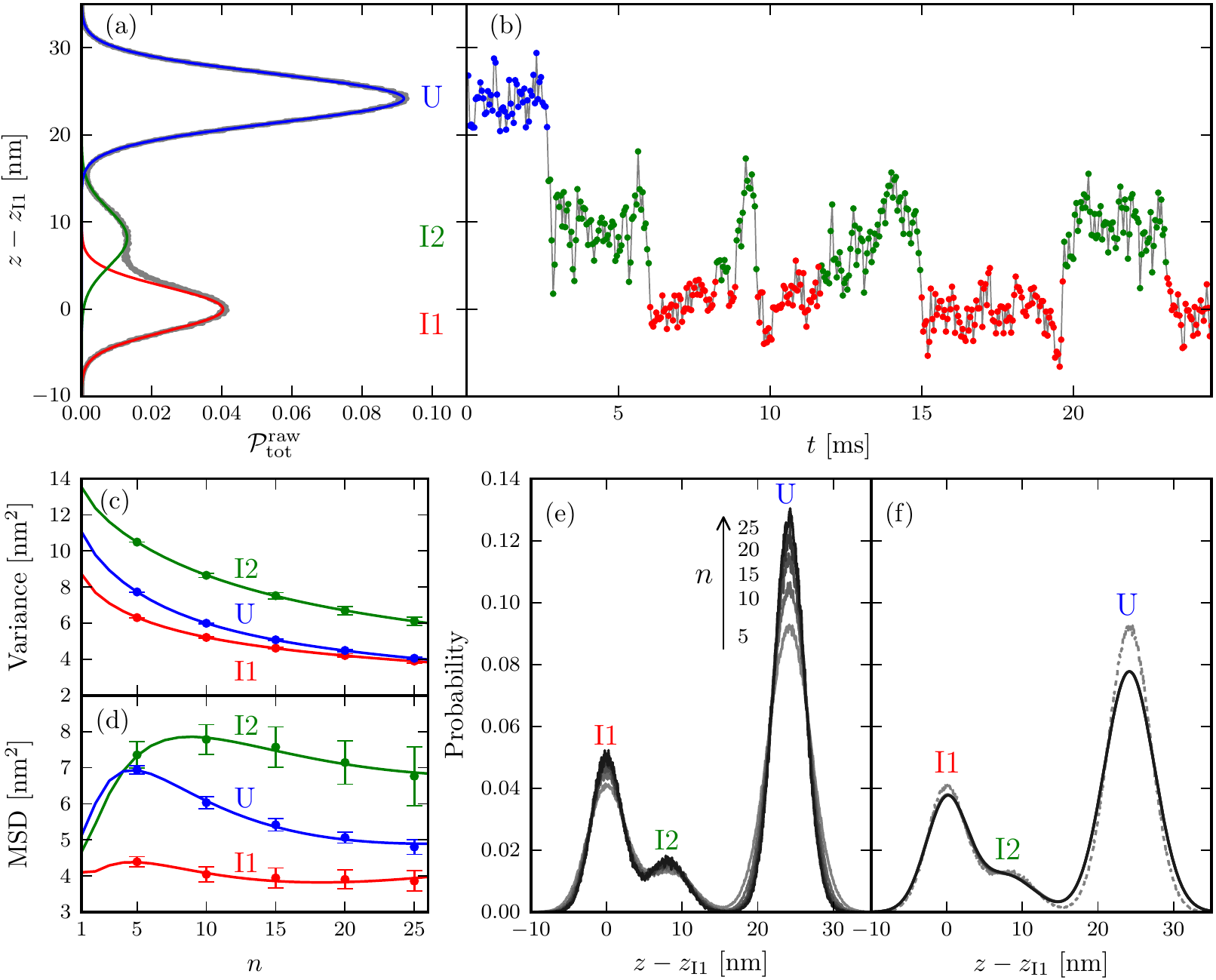}
\caption{FBS analysis of a GCN4 leucine zipper experiment (trajectory
  1, with parameters given in Table~S1).  (a) The
  probability distribution of the bead-bead separation from the raw
  experimental data, ${\cal P}^\text{raw}_\text{tot}$ (gray), and the
  decomposition into individual Gaussian peaks corresponding to the
  I1 (red), I2 (green), and U (blue) states.  Distances are
  measured with respect to $z_\text{I1}$, the position of the I1
  peak.  (b) A sample time series fragment from the experiment, with
  the individual data points colored according to their assignment to
  the I1 (red), I2 (green), and U (blue) states by hidden Markov
  model analysis.  (c) For the raw time series filtered by averaging
  together every $n$ data points, the variance $C_\text{raw,$n$}(0)$
  as a function of $n$.  The time series corresponding to each state
  is analyzed separately, and plotted as points in distinct colors,
  with bars denoting standard error due to finite sampling.  The solid
  curves are the FBS theoretical fits to $C_\text{raw,$n$}(0)$
  [Eq.~\eqref{f5}]. Best-fit FBS parameters are listed in
  Table~S2. (d) Analogous to (c), except showing the MSD function
  $\Delta_\text{raw,$n$}(n\tau_s)$ for consecutive pairs of points in
  the averaged time series.  (e) The raw distributions ${\cal
    P}^\text{raw,$n$}_\text{tot}$ for the averaged time series at
  $n=5$, 10, 15, 20, 25.  (f) Solid curves: the distribution ${\cal
    P}_\text{tot}$ estimated by applying the appropriate FBS
  correction to the raw distributions in (e).  There are five curves,
  but due to overlapping they appear as one.  Dashed curve: the raw
  distribution ${\cal P}^\text{raw,$n$}_\text{tot}$ for $n=5$.}\label{riefA_avg}
\end{figure}

The final test is on the GCN4 leucine zipper experimental time series
(trajectory 1, with parameters described in Table~S1).
As mentioned earlier, here we only can construct averaged data sets
for $n=5,10,15,\ldots$: the $n=1$ trajectory, at the original $\tau_s
= 10$ $\mu$s sampling interval, is not available.  Despite this
limitation, the FBS scaling analysis works nicely, with results
summarized in Fig.~\ref{riefA_avg} and Table~S2.  We took advantage of
the fact that the handle-only data sets, collected with the same
optical tweezer setup as the leucine zipper (except with no protein),
had direct information about $n=1$ timescales, and thus probed higher
frequencies than were accessible in the leucine zipper data.  Since
going to higher frequencies gives us better estimates of the
background noise, we set the noise amplitude $\nu$ in the leucine
zipper case to the best-fit value from the handle-only analysis.  All
other FBS parameters were fit individually for each state (I1,
I2, and U) in the leucine zipper distribution.  With FBS
corrections, ${\cal P}^\text{raw,$n$}_\text{tot}$ for $n$ going up to
25 (effective bandwidths as low as 4 kHz) all collapse onto a single
estimate of the true ${\cal P}_\text{tot}$.

\section{Estimating uncertainties in the free energy reconstruction}\label{sec:err}

The free energy reconstruction is only as good as the data on which it
is based: the recorded time series which is the input, and the
information about the apparatus which is used to analyze the time
series and predict the intrinsic landscape.  Both of these are subject
to uncertainties, which will propagate into the final result.  Let us
first consider statistical uncertainties due to the finite length of
the trajectories from which the input probability distribution ${\cal
  P}^\text{raw}_\text{tot}$ is determined.  One of the advantages of
the double optical trap setup is that it is exceptionally stable,
allowing for data collection over periods $> 100$ s.  In the case of
the leucine zipper, the slowest transition (from U to I2) occurs on
timescales of $0.4 - 0.6$ s in the force range of interest, so even a
single trajectory contains $\sim{\cal O}(10^2)$ of the rarest observed
conformational changes.

\begin{figure}[t]
\centering\includegraphics*[scale=1]{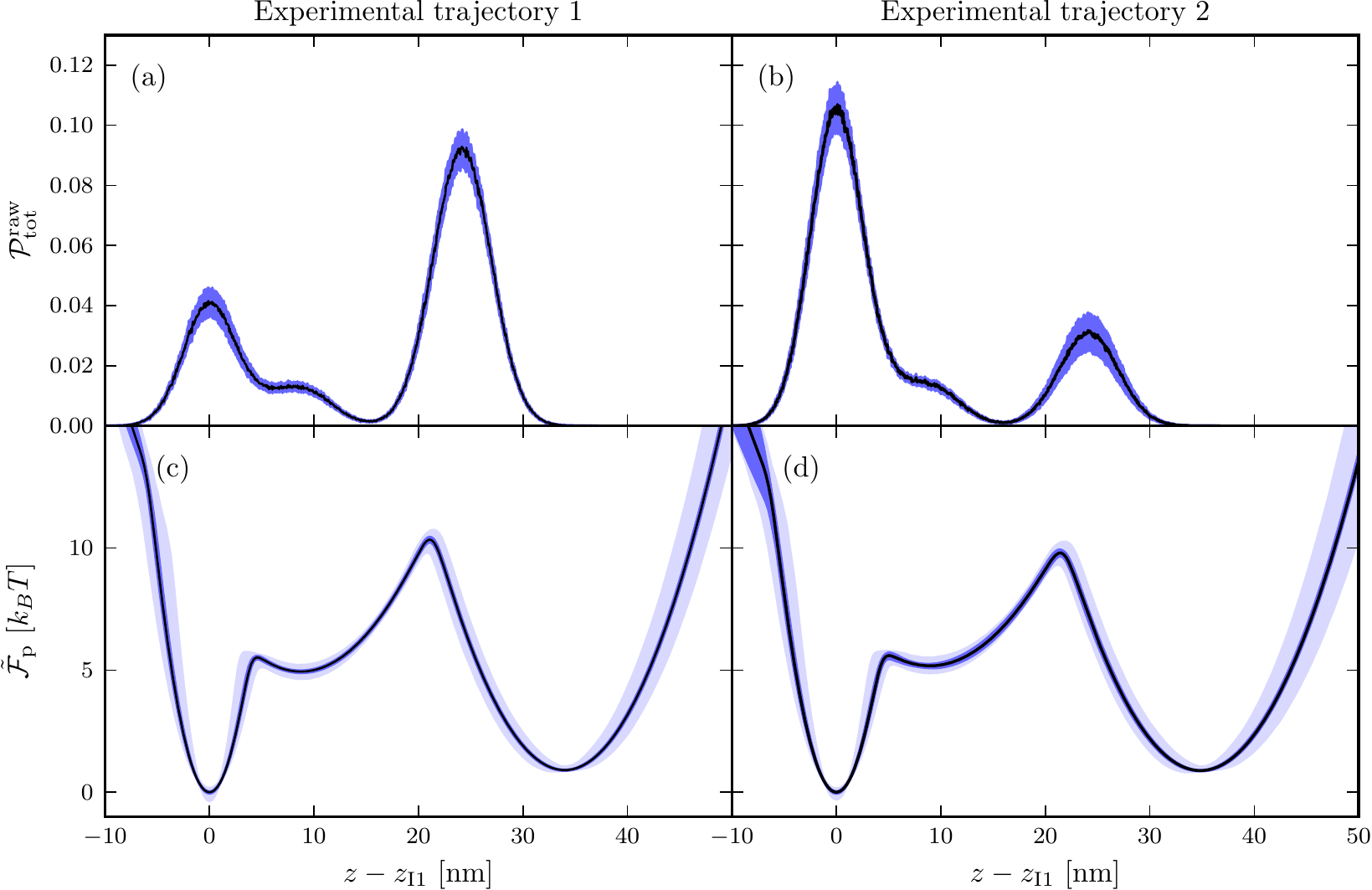}
\caption{Estimation of uncertainty in the free energy reconstruction,
  as discussed in Sec.~\ref{sec:err}.  (a-b) Probability distributions
  ${\cal P}^\text{raw}_\text{tot}$ of the raw time series for total
  bead-bead separation (black curves) collected during two
  experimental runs (left/right columns; see Table S1 for
  parameters).  Distances are measured with respect to $z_\text{I1}$, the
  position of the I1 peak.  The dark blue band corresponds to the
  standard error (68\% confidence interval) for each point in the
  distribution, due to finite sampling.  (c-d) The corresponding
  intrinsic protein free energy $\tilde{\cal F}_\text{p}$ (black
  curves), as calculated using the procedure described in the main
  text.  The free energies are in the constant force ensemble, at the
  mid-point force value $F_0$ where the probability of being in states
  I1 and U is equal ($F_0 = 12.3\pm 0.9$ pN for run 1, $12.1\pm
  0.9$ pN for run 2).  The dark blue band represents standard error
  (68\% confidence interval) including just the uncertainty due to
  finite sampling; the wider light blue band is the standard error
  including all sources of uncertainty (sampling and apparatus
  parameters).}\label{err}
\end{figure}

Thus the distribution ${\cal P}^\text{raw}_\text{tot}$ has small
statistical uncertainties.  To quantitatively estimate the error, we
use a block bootstrap method~\cite{Efron79,Politis94} in the following
manner: the trajectory is divided into blocks of length larger than
the longest autocorrelation time, and a synthetic data set of the same
length is generated by sampling with replacement from this set of
blocks.  Using a large number of synthetic data sets ($>500$) we
can determine confidence intervals for each point in the ${\cal
  P}^\text{raw}_\text{tot}$ distribution.  The number of blocks is
varied until convergence is achieved in the error estimate.  The
results are shown in Fig.~\ref{err}(a-b) for two leucine zipper
experimental trajectories (parameters as in Table S1).
The ${\cal P}^\text{raw}_\text{tot}$ distributions (black curves) are
surrounded by dark blue bands which represent the 68\% confidence
interval, or standard error margin.  The median error in the $z$ range
where ${\cal P}^\text{raw}_\text{tot} > 10^{-6}$, is 10\% and 19\%
respectively for the two trajectories.

In reconstructing the intrinsic free energy landscape $\tilde{\cal
  F}_\text{p}$, this statistical error is compounded by uncertainties
in all the apparatus parameters that are used in the analysis: bead
radii, trap strengths, handle properties, as listed in Table S1, as
well as uncertainties in the FBS parameters used to correct the raw
distributions (Table~S2).  We perform a Monte Carlo error estimate, by
sampling from Gaussian distributions of these parameters with standard
deviations given by the uncertainties, and for each parameter set
performing the complete free energy reconstruction on the entire
ensemble of synthetic data sets generated by the block bootstrap.  In
order to analyze the shape differences among the reconstructed
landscapes, every $\tilde{\cal F}_\text{p}$ is projected to the
mid-point value of $F_0$ where the probabilities of states I1 and U
are equal.  ($F_0 = 12.3\pm 0.9$ pN and $12.1 \pm 0.9$ pN from
trajectories 1 and 2 respectively.)  Though computationally intensive,
this procedure allows us to estimate 68\% confidence intervals for
$\tilde{\cal F}_\text{p}$ shown as light blue bands for the two
trajectories in Fig.~\ref{err}(c-d).  For comparison, if one assumed
no uncertainty in the apparatus parameters, one would get the much
narrower dark blue bands, representing just the error in $\tilde{\cal
  F}_\text{p}$ from the finite sampling of ${\cal
  P}^\text{raw}_\text{tot}$.  Clearly, the uncertainties in the
apparatus parameters are the predominant source of error.

With both apparatus and sampling uncertainties included, the median
standard error over the $z$ range where $\tilde{\cal F}_\text{p} <
-k_B T \ln (10^{-6})\approx 14 k_BT$ is $10\%$ in both trajectories.
This corresponds to $\approx 0.4 k_B T$ deviations in the shape of the
landscape.  The median difference between $\tilde{\cal F}_\text{p}$
estimated from the two trajectories in this range is $0.3 k_B T$, and
hence our free energy analysis gives a consistent result, within
standard error, between the two different experimental runs.

\singlespacing

\SkipTocEntry

\vspace{1em}
\end{widetext}

\end{document}